\documentclass[journal]{IEEEtran}

\newcommand{\etal}{\textit{et al}. }

\usepackage[utf8]{inputenc} 
\usepackage[T1]{fontenc}    
\usepackage{hyperref}       
\usepackage{url}            
\usepackage{booktabs}       
\usepackage{amsfonts}       
\usepackage{nicefrac}       
\usepackage{microtype}      
\usepackage{amsmath}
\usepackage{bbm}
\usepackage{array}
\usepackage{wrapfig}
\usepackage{arydshln}
\usepackage[export]{adjustbox}
\usepackage{xparse}
\usepackage{multirow}
\usepackage{cite}
\usepackage[caption=false, font=footnotesize]{subfig}
\usepackage{cprotect}

\usepackage{tikz}
\let\svtikzpicture\tikzpicture
\def\tikzpicture{\noindent\svtikzpicture}
\usepackage{lipsum,adjustbox,cleveref}
\usetikzlibrary{calc,positioning}
\usetikzlibrary{shapes,arrows}

\usepackage{algpseudocode}
\usepackage{algorithm}
\algnewcommand\algorithmicforeach{\textbf{for}}
\algdef{S}[FOR]{ForEach}[1]{\algorithmicforeach\ #1\ \algorithmicdo}

\newcommand{\argmin}{\operatornamewithlimits{argmin}}

\NewDocumentCommand{\ShowInline}{v}{%
#1%
}

\makeatletter
\long\def\@makecaption#1#2{\ifx\@captype\@IEEEtablestring%
\footnotesize\begin{center}{\normalfont\footnotesize #1}\\
{\normalfont\footnotesize\scshape #2}\end{center}%
\@IEEEtablecaptionsepspace
\else
\@IEEEfigurecaptionsepspace
\setbox\@tempboxa\hbox{\normalfont\footnotesize {#1.}~~ #2}%
\ifdim \wd\@tempboxa >\hsize%
\setbox\@tempboxa\hbox{\normalfont\footnotesize {#1.}~~ }%
\parbox[t]{\hsize}{\normalfont\footnotesize \noindent\unhbox\@tempboxa#2}%
\else
\hbox to\hsize{\normalfont\footnotesize\hfil\box\@tempboxa\hfil}\fi\fi}
\makeatother
\title{CONVIQT: Contrastive Video Quality Estimator}

\author{Pavan C. Madhusudana, Neil Birkbeck, Yilin Wang,  Balu Adsumilli and Alan C. Bovik 
	\thanks{P. C. Madhusudana and A. C. Bovik are with the Department of Electrical and
Computer Engineering, University of Texas at Austin, Austin, TX, USA (e-mail:
pavancm@utexas.edu; bovik@ece.utexas.edu). Neil Birkbeck, Yilin Wang
and Balu Adsumilli are with Google Inc. (e-mail: birkbeck@google.com; yilin@google.com; badsumilli@google.com).}}

\begin{document}

\maketitle

\begin{abstract}
Perceptual video quality assessment (VQA) is an integral component of many streaming and video sharing platforms. Here we consider the problem of learning perceptually relevant video quality representations in a self-supervised manner. Distortion type identification and degradation level determination is employed as an auxiliary task to train a deep learning model containing a deep Convolutional Neural Network (CNN) that extracts spatial features, as well as a recurrent unit that captures temporal information. The model is trained using a contrastive loss and we therefore refer to this training framework and resulting model as \textbf{CON}trastive \textbf{VI}deo \textbf{Q}uality Estima\textbf{T}or (CONVIQT). During testing, the weights of the trained model are frozen, and a linear regressor maps the learned features to quality scores in a no-reference (NR) setting. We conduct comprehensive evaluations of the proposed model on multiple VQA databases by analyzing the correlations between model predictions and ground-truth quality ratings, and achieve competitive performance when compared to state-of-the-art NR-VQA models, even though it is not trained on those databases. Our ablation experiments demonstrate that the learned representations are highly robust and generalize well across synthetic and realistic distortions. Our results indicate that compelling representations with perceptual bearing can be obtained using self-supervised learning. The implementations used in this work have been made available at \url{https://github.com/pavancm/CONVIQT}. 
\end{abstract}


\section{Introduction}
\IEEEPARstart{T}{he} smartphone revolution has led to an explosion of video consumption over the internet. A recent report by Cisco \cite{cisco} estimates that more than 82\% of internet traffic will be dominated by online videos. Popular video sharing and streaming platforms such as YouTube, Facebook, Netflix, and Amazon Prime Video are accessed by hundreds of millions of people over the globe, and more than one billion hours of video are watched every day. Given the central nature of videos as a primary medium of communication, it is important that these platforms be able to monitor and control perceptual video quality and to deliver better consumer experiences. The problem of objectively quantifying and estimating perceptual video quality is referred to as Video Quality Assessment (VQA). Although subjective VQA, where quality judgments are obtained by human observers, provides the most accurate and reliable quality estimates, it is generally impractical, and does not scale with the number of videos being assessed. This necessitates the need for developing objective VQA models that can reliably predict the responses of human observers on the task of quality estimation. The goal of VQA models is to obtain predictions that correlate well with subjective opinions. No-reference (NR, or blind) VQA models are constrained to predict the quality of distorted videos without any knowledge of any high quality reference or of the artifacts that afflict them. The easy availability of affordable video capture devices coupled with surge in smartphone usage has led to massive amount of User Generated Content (UGC) videos being continuously uploaded to social media sites. Since UGC videos vary considerably in the amount of distortions that come from capture, and since they are usually compressed introducing further distortions. To achieve efficient handling of this video traffic it is necessary to be able to objectively quantify perceptual video quality and to guide subsequent processing tasks such as further compression \cite{yu2019predicting}, enhancement, and optimization of bandwidth versus video quality.

Due to the practical significance of NR-VQA, a diverse set of VQA databases and objective models have been created over the last decade. A VQA database contains a set of distorted videos, along with corresponding human opinion scores of them obtained either in a laboratory or a crowd-sourced environment. Earlier databases, such as LIVE-VQA \cite{seshadrinathan2010study}, CSIQ-VQA \cite{vu2014vis3}, EPFLPoliMI \cite{de2010h}, and MCL-V \cite{lin2015mcl} datasets each contain a small set of pristine, high quality source contents corrupted with compression and transmission artifacts. Each source video was synthetically degraded by only one or two types of distortions applied in a controlled manner. Although these databases were influential in improving video quality models, they do not represent the reality of pervasive UGC videos that are commonly corrupted by innumerable combinations of multiple distortions. Because of this, several large scale databases that contain authentic UGC distortions are now available such as LIVE-VQC \cite{sinno2018large}, KoNViD \cite{hosu2017konstanz}, YouTube-UGC \cite{wang2019youtube} and LSVQ \cite{ying2021patch}. The UGC-VQA problem is particularly challenging due to the large variety of distortions that can occur, as the influence of video content on perceived quality. Advancements in capture and display technologies has led to the widespread adoption of videos of 4K/ultra high definition (UHD) resolution, high dynamic range (HDR) and high frame rate (HFR). This has led to the creation of datasets like AVT-VQDB-UHD-1 \cite{rao2019avt}, ETRI-LIVE STSVQ \cite{lee2021subjective}, BVI-HFR \cite{mackin2018study} and LIVE-YT-HFR \cite{madhusudana2021liveythfr} which now make it possible to analyze the perceptual implications of these new developments. From an NR algorithm design standpoint, having a single reliable model that is able to accurately quantify both synthetic and authentic artifacts would be quite advantageous, since it could be used in wide ranges of scenario presenting highly diverse of the distortions and combinations of them.

Popular NR-VQA models employ a two stage design, where in the first stage quality relevant features are extracted followed by a regression stage that maps the features to perceptual quality scores. Earlier techniques relied heavily on the use of Natural Scene Statistics (NSS) to derive features that are highly sensitive to visual quality. Although these models have been successful at predicting the perception of synthetically applied distortions, their performance has proven to be limited on the realistic distortions encountered in UGC videos. End-to-end trained deep learning based VQA models \cite{li2019quality,ying2021patch,chen2021learning,li2022blindly} have achieved significant improvements over these traditional methods on UGC content. These performance improvements have mainly been arrived at using transfer learning techniques, whereby a deep model is first trained on a large labeled dataset such as ImageNet \cite{russakovsky2015imagenet} or Kinetics-400 \cite{kay2017kinetics}, then later fine-tuned on human opinion scores. Transfer learning is employed since currently available VQA datasets are generally too small to train deep models from scratch. Although fine-tuning can be an effective way to obtain good performance on both synthetic and realistic distortions, it requires careful choice of hyperparameters, which may change on different VQA databases. 

A recent still picture (spatial) model called CONTRIQUE \cite{madhusudana2021image} employs a self-supervised training approach, where a deep model was trained to distinguish between different types of image artifacts and also to determine the degree of perceptual degradation present. The main advantage of this approach is that it does not require human opinion scores of visual quality for training, making model training feasible on unlabeled image databases. The performance of CONTRIQUE was shown to be highly competitive on multiple image quality datasets comprising both synthetic and authentic artifacts. Motivated by the successes of CONTRIQUE, here we extend the idea to video quality prediction, which is a much harder problem of higher dimensionality. We refer to the new model as the \textbf{CON}trastive \textbf{VI}deo \textbf{Q}uality Estima\textbf{T}or (CONVIQT). Notable attributes of CONVIQT are as follows:

\begin{enumerate}
    \item Similar to CONTRIQUE, the CONVIQT model is first trained from scratch to solve the auxiliary tasks of distortion type and degradation level discrimination. Training is done on an unlabeled video dataset containing a mix of synthetic and realistic video distortions, using a contrastive loss function.
    \item The CONVIQT architecture includes the existing CONTRIQUE model which is used to capture spatial attributes of video quality. The weights of the CONTRIQUE model are frozen and not modified during the training of CONVIQT.
    \item In addition, recurrent unit processes the spatial features to extract temporal quality information. The recurrent unit is the only trainable component in the CONVIQT pipeline.
    \item During evaluation of the CONVIQT representations, all of the weights are frozen and a linear regressor on each VQA dataset learns to map CONVIQT features quality scores. The quality predictions produced by CONVIQT obtain competitive performance as compared to other state-of-the-art (SOTA) VQA models across multiple databases even though those models are trained on them. This is achieved without any additional fine-tuning of the backbone deep learning model.
    \item The CONVIQT framework is simple but effective, and generalizes well across synthetic and realistic distortions.
\end{enumerate}
        
The rest of the paper is organized as follows: In Section \ref{sec:prior_work} we discuss related work on VQA and self-supervised learning. Section \ref{sec:Method} provides a detailed description of the design of CONVIQT. Section \ref{sec:experiments} analyzes and compares various experimental results obtained with CONVIQT, and Section \ref{sec:conclusion} provides concluding remarks.

\section{Related Work}
\label{sec:prior_work}
Next we discuss related work on NR-VQA and self-supervised learning.

\tikzstyle{block} = [draw, fill=blue!20, rounded corners, text centered,  minimum height=3em, minimum width=4em]
\tikzstyle{block2} = [draw, rounded corners, text centered,  minimum height=3em, minimum width=4em]
\tikzstyle{circle1} = [draw, fill=blue!20, circle, text centered,  minimum size=0.5em]
\tikzstyle{input} = [coordinate]
\tikzstyle{output} = [coordinate]
\tikzstyle{pinstyle} = [pin edge={to-,thin,black}]
\pgfdeclarelayer{background}
\pgfdeclarelayer{foreground}
\pgfsetlayers{background,main,foreground}

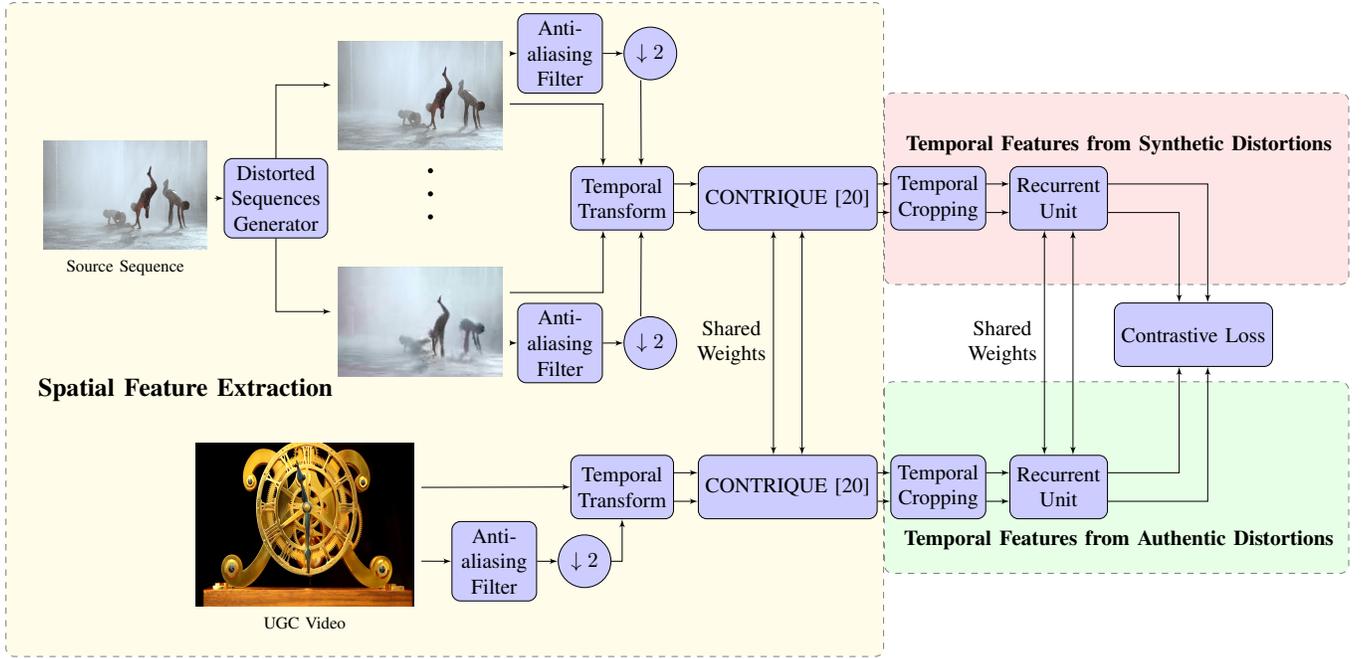
\begin {figure*}[t]
\captionsetup[subfigure]{justification = centering,labelformat=empty,position=bottom}
\begin{minipage}[b]{1\linewidth}
	\centering
	\resizebox{1\textwidth}{!}{%
		\begin{tikzpicture}[auto, node distance=2cm,>=latex']
		
		\node (nat_im) [left = 1.5cm,align=center]{\subfloat[Source Sequence]{\includegraphics[width = 0.15\linewidth, height = 0.1\linewidth]{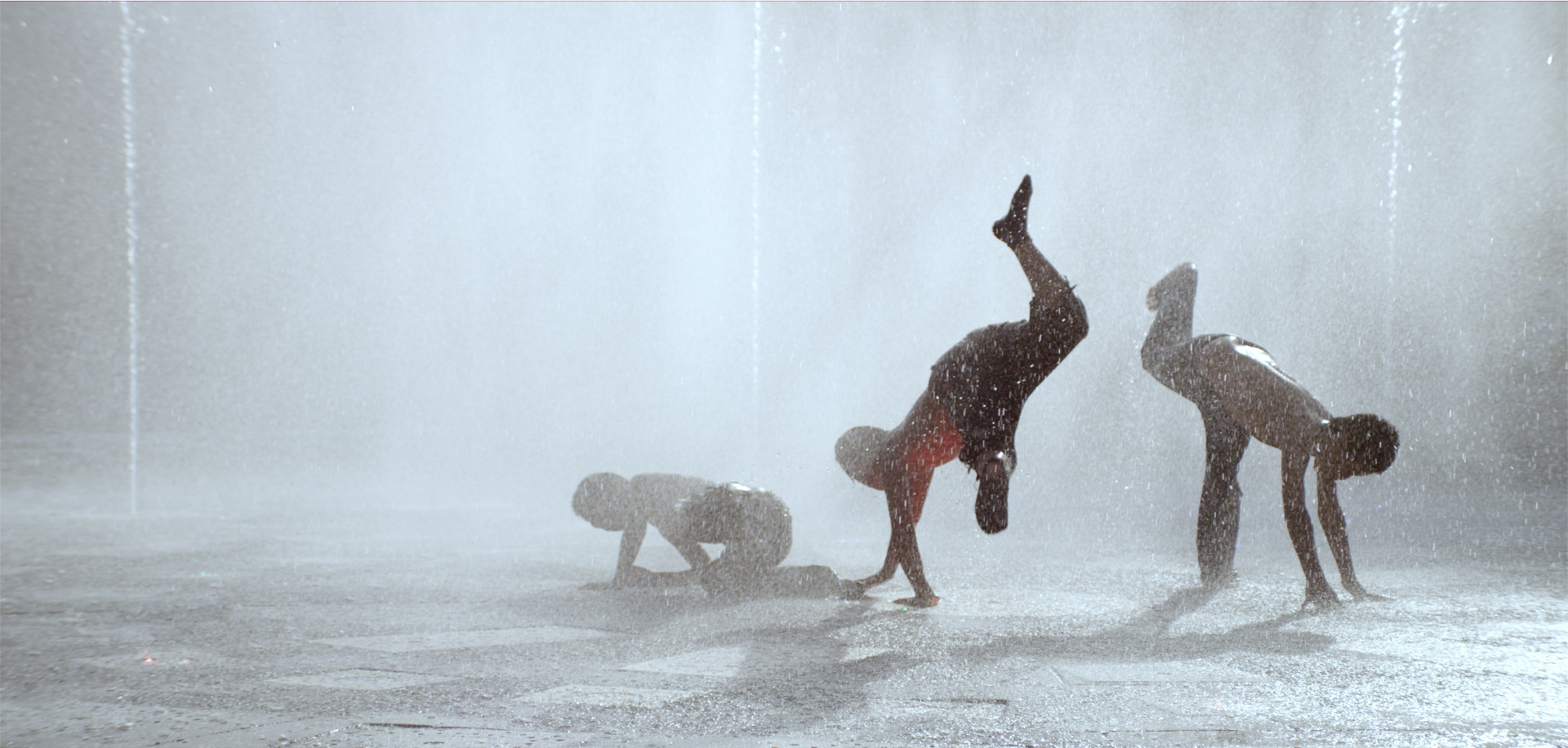}}};
		
		\node (syn_dist) [block, right of=nat_im,node distance=2.5cm, align=center]{Distorted \\ Sequences \\ Generator};
		
		\node (syn1) [above right = 0.01cm and 0.05cm of syn_dist, align=center]{\subfloat{\includegraphics[width=0.15\linewidth,height = 0.1\linewidth]{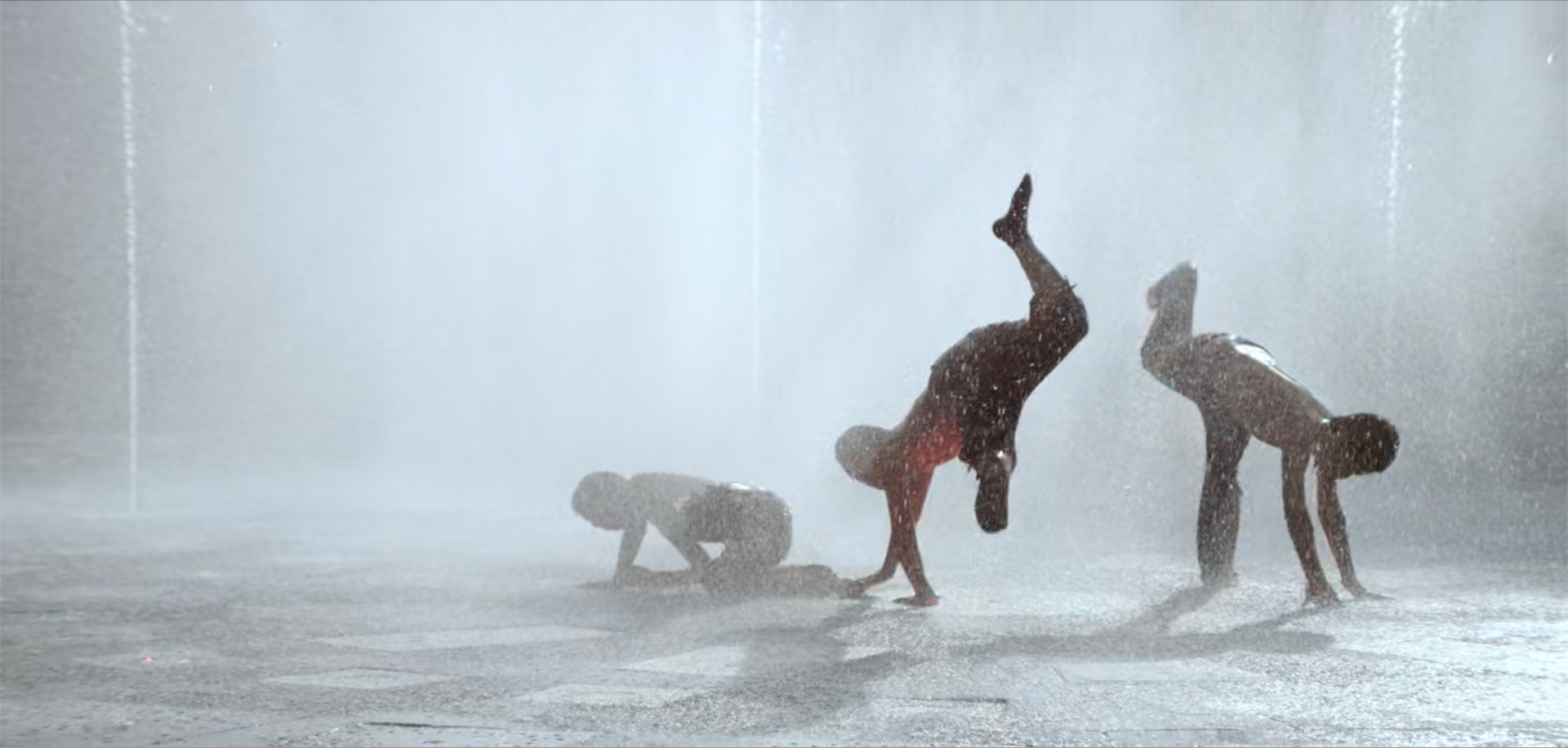}}};
		
		\node (syn2) [below right = 0.01cm and 0.05cm of syn_dist, align=center]{\subfloat{\includegraphics[width=0.15\linewidth,height = 0.1\linewidth]{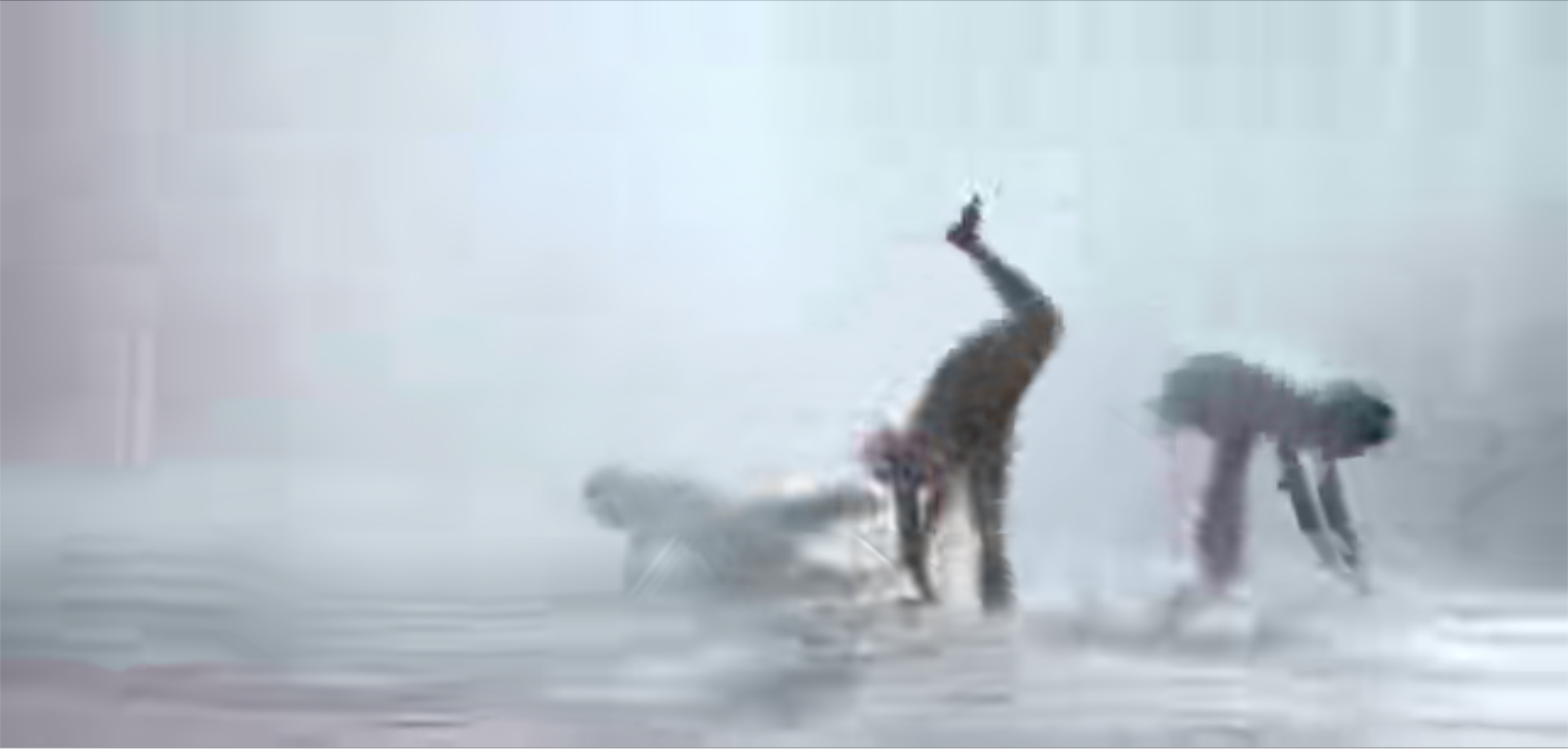}}};
		
		\node (r_patch1) [block, right of=syn_dist, node distance = 5.75cm, align=center] {Temporal \\ Transform};
		\node (alias1)[block, below right = -1.35cm and 0.125cm of syn2, align=center] {Anti-\\aliasing\\Filter};
		\node (alias12)[block, above right = -1.35cm and 0.125cm of syn1, align=center] {Anti-\\aliasing\\Filter};
		\node (ds1) [circle1, right of=alias1, node distance = 1.5cm, align=center] {$\downarrow 2$};
		\node (ds12) [circle1, right of=alias12, node distance = 1.5cm, align=center] {$\downarrow 2$};
		\node (enc1) [block, right of=r_patch1, node distance = 2.75cm, align=center] {CONTRIQUE \cite{madhusudana2021image}};
		\node (tcrop1) [block, right of=enc1, node distance = 2.5cm, align=center] {Temporal \\ Cropping};
		\node (local1) [block, right of=tcrop1, node distance = 2cm, align=center] {Recurrent \\ Unit};
		
		\node (auth_im) [below left = 0.5cm and -1.5cm of syn2, node distance = 4.5cm, align=center] {\subfloat[UGC Video]{\includegraphics[width = 0.2\linewidth,height = 0.15\linewidth]{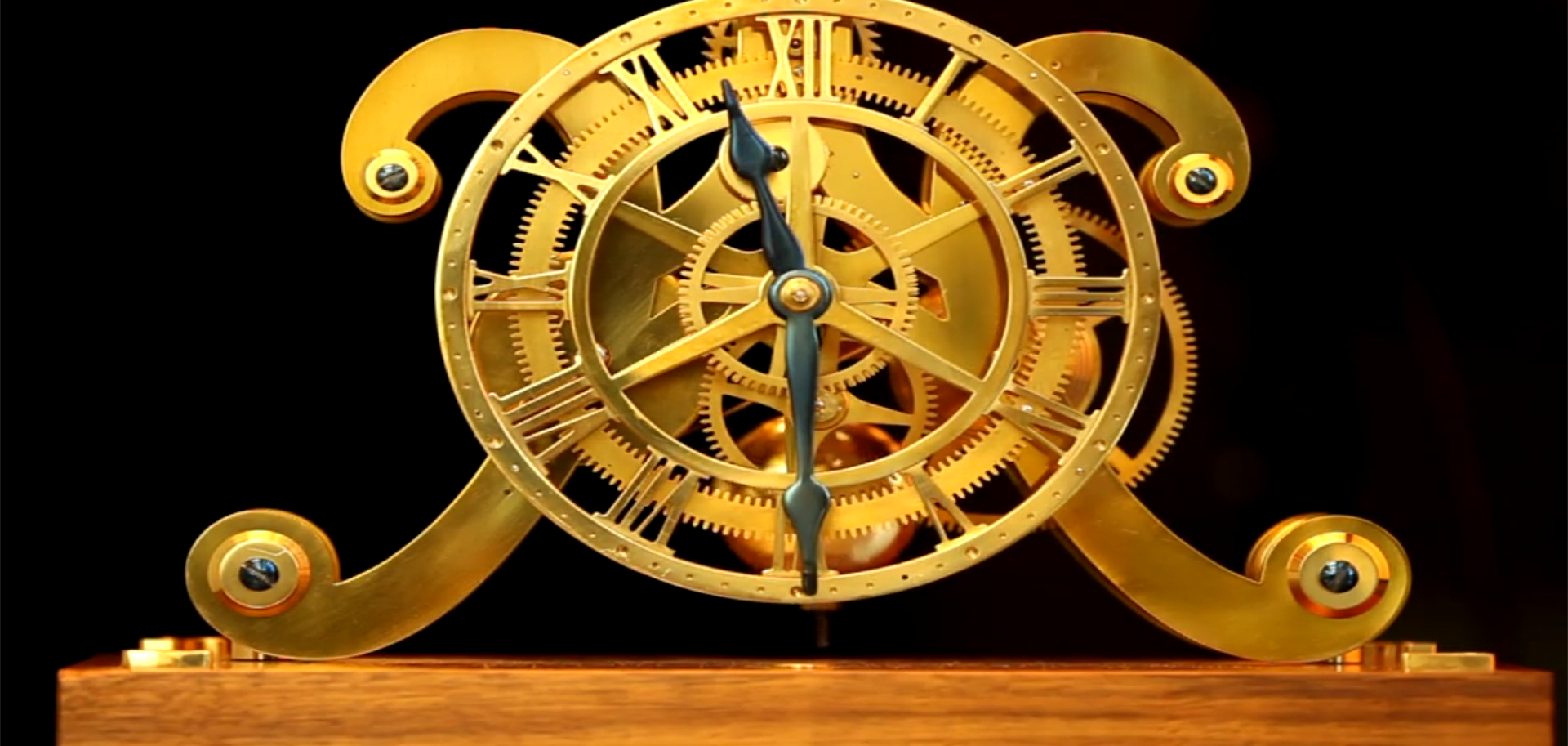}}};
		\node (r_patch2) [block, below of=r_patch1, node distance = 4.8cm, align=center] {Temporal \\ Transform};
		\node (alias2)[block, below right = -2cm and 0.5cm of auth_im, align=center] {Anti-\\aliasing\\Filter};
		\node (ds2) [circle1, right of=alias2, node distance = 1.5cm, align=center] {$\downarrow 2$};
		
		\node (enc2) [block, right of=r_patch2, node distance = 2.75cm, align=center] {CONTRIQUE \cite{madhusudana2021image}};
		\node (tcrop2) [block, right of=enc2, node distance = 2.5cm, align=center] {Temporal \\ Cropping};
		\node (local2) [block, right of=tcrop2, node distance = 2cm, align=center] {Recurrent \\ Unit};
		
		\node (c_loss) [block, below right = 1.2cm and 0.1cm of local1, align=center] {Contrastive Loss};
		
		\path (syn1.south) -- (syn2.north) node[font=\Huge, midway, sloped]{$\dots$};
		\draw [->] (nat_im) -- node{}(syn_dist);
		\draw [->] (syn_dist) |- node{}(syn1);
		\draw [->] (syn_dist) |- node{}(syn2);
		\draw [->] ([yshift=2ex]syn2.east) -| node{}([xshift=-2ex]r_patch1.south);
		\draw [->] ([yshift=-3.3ex]syn2.east) -- node{}(alias1.west);
		\draw [->] (alias1.east) -- node{}(ds1.west);
		\draw [->] ([yshift=-2ex]syn1.east) -| node{}([xshift=-2ex]r_patch1.north);
		\draw [->] ([yshift=3.3ex]syn1.east) -- node{}(alias12.west);
		\draw [->] (alias12.east) -- node{}(ds12.west);
		\draw [->] ([yshift=4.25ex]auth_im.east) -- node{}(r_patch2.west);
		\draw [->] ([yshift=-3.5ex]auth_im.east) -- node{}(alias2.west);
		\draw [->] (alias2.east) -- node{}(ds2.west);
		
		\draw [->] ([xshift=-1ex]ds1.north) -- node{}([xshift=2ex]r_patch1.south);
		\draw [->] ([xshift=-1ex]ds12.south) -- node{}([xshift=2ex]r_patch1.north);
		\draw [->] (ds2) -| node{}(r_patch2.south);
		\draw [->] ([yshift=-1.5ex]r_patch1.east) -- node{}([yshift=-1.5ex]enc1.west);
		\draw [->] ([yshift=1.5ex]r_patch1.east) -- node{}([yshift=1.5ex]enc1.west);
		\draw [->] ([yshift=-1.5ex]r_patch2.east) -- node{}([yshift=-1.5ex]enc2.west);
		\draw [->] ([yshift=1.5ex]r_patch2.east) -- node{}([yshift=1.5ex]enc2.west);
		\draw [->] ([yshift=-1.5ex]enc1.east) -- node{}([yshift=-1.5ex]tcrop1.west);
		\draw [->] ([yshift=1.5ex]enc1.east) -- node{}([yshift=1.5ex]tcrop1.west);
		\draw [->] ([yshift=-1.5ex]tcrop1.east) -- node{}([yshift=-1.5ex]local1.west);
		\draw [->] ([yshift=1.5ex]tcrop1.east) -- node{}([yshift=1.5ex]local1.west);
		\draw [->] ([yshift=-1.5ex]enc2.east) -- node{}([yshift=-1.5ex]tcrop2.west);
		\draw [->] ([yshift=1.5ex]enc2.east) -- node{}([yshift=1.5ex]tcrop2.west);
		\draw [->] ([yshift=-1.5ex]tcrop2.east) -- node{}([yshift=-1.5ex]local2.west);
		\draw [->] ([yshift=1.5ex]tcrop2.east) -- node{}([yshift=1.5ex]local2.west);
		\draw [->] ([yshift=-1.5ex]local1.east) -| node{}([xshift=-1.5ex]c_loss.north);
		\draw [->] ([yshift=1.5ex]local1.east) -| node{}([xshift=1.5ex]c_loss.north);
		\draw [->] ([yshift=-1.5ex]local2.east) -| node{}([xshift=1.5ex]c_loss.south);
		\draw [->] ([yshift=1.5ex]local2.east) -| node{}([xshift=-1.5ex]c_loss.south);
		
		\draw[<->] ([xshift=-1.5ex]enc1.south) -- node[left, align=center]{Shared \\ Weights}([xshift=-1.5ex]enc2.north);
		\draw[<->] ([xshift=1.5ex]enc1.south) -- node{}([xshift=1.5ex]enc2.north);
		\draw[<->] ([xshift=-1.5ex]local1.south) -- node[left, align=center]{Shared \\ Weights}([xshift=-1.5ex]local2.north);
		\draw[<->] ([xshift=1.5ex]local1.south) -- node{}([xshift=1.5ex]local2.north);
		
		\begin{pgfonlayer}{background}
		\path (nat_im.west)+(-0.5,3.25) node (a) {};
		\path (enc1.east |- auth_im.south)+(+0.1,-0.25) node (c) {};
		
		\path[fill=yellow!10,rounded corners, draw=black!50, dashed](a) rectangle (c);           
		
		\end{pgfonlayer}
		
		\begin{pgfonlayer}{background}
		\path (tcrop1.west)-|+(-0.1,1.75) node (d) {};
		\path (local1.east |- local1.south)+(+4,-0.9) node (e) {};
		
		\path[fill=red!10,rounded corners, draw=black!50, dashed](d) rectangle (e);           
		
		\end{pgfonlayer}
		
		\begin{pgfonlayer}{background}
		\path (tcrop2.west)-|+(-0.1,1.75) node (d) {};
		\path (local2.east |- local2.south)+(+4,-0.9) node (e) {};
		
		\path[fill=green!10,rounded corners, draw=black!50, dashed](d) rectangle (e);           
		
		\end{pgfonlayer}
		
		\path (local1.north) +(1,0.35) 
		node (asrs) {\textbf{Temporal Features from Synthetic Distortions}};
		
		\path (local2.south) +(1.0,-0.35) 
		node (asrs) {\textbf{Temporal Features from Authentic Distortions}};
		
		\path (nat_im.south) +(1.0,-1.75) 
		node (asrs) {\textbf{\large{Spatial Feature Extraction}}};
		\end{tikzpicture}
	}
\end{minipage}
\centering\caption{Illustration of CONVIQT training pipeline.}
\label{fig:overview_video}
\end{figure*}

\subsection{NR-VQA Models}
Objective VQA models can be broadly categorized based on the availability of reference information: Full-Reference (FR), Reduced-Reference (RR) and No-Reference (NR). FR/RR VQA \cite{VMAF2016,bampis2018spatiotemporal,madhusudana2021st,madhusudana2022making} models employ high quality pristine reference videos against which the perceptual fidelity of distorted videos are compared when determining quality, while NR models make judgments without any reference information. Here our primary focus will be on NR-VQA methods.

\tikzstyle{block} = [draw, fill=blue!20, rounded corners, text centered,  minimum height=3em, minimum width=4em]
\tikzstyle{block2} = [draw, rounded corners, text centered,  minimum height=3em, minimum width=4em]
\tikzstyle{circle1} = [draw, fill=blue!20, circle, text centered,  minimum size=0.5em]
\tikzstyle{input} = [coordinate]
\tikzstyle{output} = [coordinate]
\tikzstyle{pinstyle} = [pin edge={to-,thin,black}]
\pgfdeclarelayer{background}
\pgfdeclarelayer{foreground}
\pgfsetlayers{background,main,foreground}

\begin {figure*}[t]
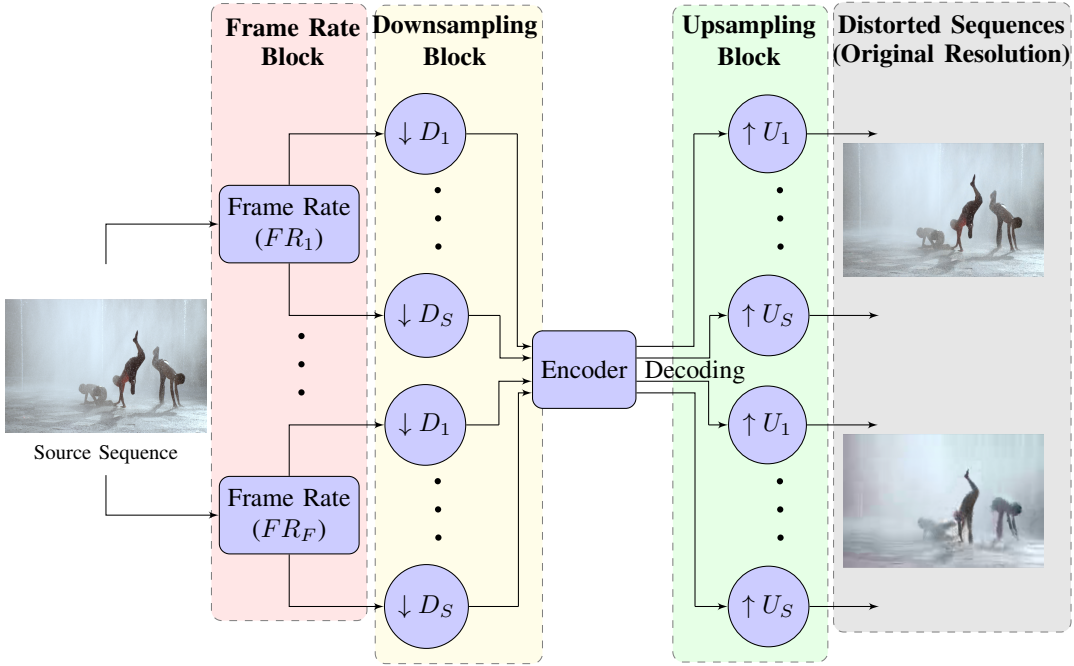

\captionsetup[subfigure]{justification = centering,labelformat=empty,position=bottom}
\begin{minipage}[b]{1\linewidth}
	\centering
	\resizebox{0.8\textwidth}{!}{%
		\begin{tikzpicture}[auto, node distance=2cm,>=latex']
		
		\node (nat_im) [left = 1.5cm,align=center]{\subfloat[Source Sequence]{\includegraphics[width = 0.15\linewidth, height = 0.1\linewidth]{Images/ref_syn.png}}};
		
		\node (fr1) [block, above right = 0.01cm and 0.05cm of nat_im, align=center]{Frame Rate \\ ($FR_1$)};
		\node (fr2) [block, below right = 0.01cm and 0.05cm of nat_im, align=center]{Frame Rate \\ ($FR_F$)};
		
		\node (ds11) [circle1, above right = 0.3cm and 0.5cm of fr1, align=center] {$\downarrow D_1$};
		\node (ds12) [circle1, below right = 0.3cm and 0.5cm of fr1, align=center] {$\downarrow D_S$};
		
		\node (ds21) [circle1, above right = 0.3cm and 0.5cm of fr2, align=center] {$\downarrow D_1$};
		\node (ds22) [circle1, below right = 0.3cm and 0.5cm of fr2, align=center] {$\downarrow D_S$};
		
		\node (dummy_d11)[input,right of=ds11,node distance = 1.25cm, align=center] {};
		\node (dummy_d12)[input,right of=ds12,node distance = 1cm, align=center] {};
		\node (dummy_d21)[input,right of=ds21,node distance = 1cm, align=center] {};
		\node (dummy_d22)[input,right of=ds22,node distance = 1.25cm, align=center] {};
		
		\node (enc) [block, right of=nat_im, node distance = 6.5cm, align=center] {Encoder};
		
		\node (us11) [circle1, right of=ds11, node distance = 4.65cm, align=center] {$\uparrow U_1$};
		\node (us12) [circle1, right of=ds12, node distance = 4.65cm, align=center] {$\uparrow U_S$};
		
		\node (us21) [circle1, right of=ds21, node distance = 4.65cm, align=center] {$\uparrow U_1$};
		\node (us22) [circle1, right of=ds22, node distance = 4.65cm, align=center] {$\uparrow U_S$};
		
		\node (dummy_u11)[input,right of=ds11,node distance = 3.65cm, align=center] {};
		\node (dummy_u12)[input,right of=ds12,node distance = 3.85cm, align=center] {};
		\node (dummy_u21)[input,right of=ds21,node distance = 3.85cm, align=center] {};
		\node (dummy_u22)[input,right of=ds22,node distance = 3.65cm, align=center] {};
		
		\node (syn1) [above right = 0.01cm and 0.5cm of us12, align=center]{\subfloat{\includegraphics[width=0.15\linewidth,height = 0.1\linewidth]{Images/dist_syn1.png}}};
		
		\node (syn2) [above right = 0.01cm and 0.5cm of us22, align=center]{\subfloat{\includegraphics[width=0.15\linewidth,height = 0.1\linewidth]{Images/dist_syn2.png}}};
		
		\node (dummy11)[input,right of=us11,node distance = 1.5cm, align=center] {};
		\node (dummy12)[input,right of=us12,node distance = 1.5cm, align=center] {};
		\node (dummy21)[input,right of=us21,node distance = 1.5cm, align=center] {};
		\node (dummy22)[input,right of=us22,node distance = 1.5cm, align=center] {};
		
		\path (fr1.south) -- (fr2.north) node[font=\Huge, midway, sloped]{$\dots$};
		\path (ds11.south) -- (ds12.north) node[font=\Huge, midway, sloped]{$\dots$};
		\path (ds21.south) -- (ds22.north) node[font=\Huge, midway, sloped]{$\dots$};
		
		\path (us11.south) -- (us12.north) node[font=\Huge, midway, sloped]{$\dots$};
		\path (us21.south) -- (us22.north) node[font=\Huge, midway, sloped]{$\dots$};
		
		\draw [->] (nat_im.south) |- node{}(fr2.west);
		\draw [->] (nat_im.north) |- node{}(fr1.west);
		
		\draw [->] (fr1.south) |- node{}(ds12.west);
		\draw [->] (fr1.north) |- node{}(ds11.west);
		
		\draw [->] (fr2.south) |- node{}(ds22.west);
		\draw [->] (fr2.north) |- node{}(ds21.west);
		
		\draw [-] (ds11.east) -- node{}(dummy_d11);
		\draw [->] (dummy_d11) |- node{}([yshift=2ex]enc.west);
		
		\draw [-] (ds12.east) -- node{}(dummy_d12);
		\draw [->] (dummy_d12) |- node{}([yshift=1ex]enc.west);
		
		\draw [-] (ds21.east) -- node{}(dummy_d21);
		\draw [->] (dummy_d21) |- node{}([yshift=-1ex]enc.west);
		
		\draw [-] (ds22.east) -- node{}(dummy_d22);
		\draw [->] (dummy_d22) |- node{}([yshift=-2ex]enc.west);
		
		\draw [-] ([yshift=2ex]enc.east) -| node{}(dummy_u11);
		\draw [->] (dummy_u11) -- node{}(us11.west);
		
		\draw [-] ([yshift=1ex]enc.east) -| node{}(dummy_u12);
		\draw [->] (dummy_u12) -- node{}(us12.west);
		
		\draw [-] ([yshift=-1ex]enc.east) -| node{}(dummy_u21);
		\draw [->] (dummy_u21) -- node{}(us21.west);
		
		\draw [-] ([yshift=-2ex]enc.east) -| node[above]{Decoding}(dummy_u22);
		\draw [->] (dummy_u22) -- node{}(us22.west);
		
		\draw [->] (us11.east) -- node{}(dummy11.west);
		\draw [->] (us12.east) |- node{}(dummy12.west);
		
		\draw [->] (us21.east) -- node{}(dummy21.west);
		\draw [->] (us22.east) |- node{}(dummy22.west);
		
		\begin{pgfonlayer}{background}
		\path (fr1.west)-|+(-0.1,3) node (d) {};
		\path (fr2.east |- fr2.south)+(0.1,-0.9) node (e) {};
		
		\path[fill=red!10,rounded corners, draw=black!50, dashed](d) rectangle (e);           
		
		\end{pgfonlayer}
		
		\begin{pgfonlayer}{background}
		\path (ds11.west)-|+(-0.125,1.75) node (d) {};
		\path (ds22.east |- ds22.south)+(1,-0.2) node (e) {};
		
		\path[fill=yellow!10,rounded corners, draw=black!50, dashed](d) rectangle (e);           
		\end{pgfonlayer}
		
		\begin{pgfonlayer}{background}
		\path (us11.west)-|+(-0.75,1.75) node (d) {};
		\path (us22.east |- us22.south)+(0.25,-0.2) node (e) {};
		
		\path[fill=green!10,rounded corners, draw=black!50, dashed](d) rectangle (e);           
		\end{pgfonlayer}
		
		\begin{pgfonlayer}{background}
		\path (syn1.west)-|+(-0.01,2.65) node (d) {};
		\path (syn2.east |- syn2.south)+(0.25,-0.75) node (e) {};
		
		\path[fill=black!10,rounded corners, draw=black!50, dashed](d) rectangle (e);           
		\end{pgfonlayer}
		
		\path (fr1.north) +(0.05,2.15) 
		node (asrs) {\textbf{Frame Rate}};
		
		\path (fr1.north) +(0.05,1.75) 
		node (asrs) {\textbf{Block}};
		
		\path (fr1.north) +(2.25,2.15) 
		node (asrs) {\textbf{Downsampling}};
		
		\path (fr1.north) +(2.25,1.75) 
		node (asrs) {\textbf{Block}};
		
		\path (fr1.north) +(6.25,2.15) 
		node (asrs) {\textbf{Upsampling}};
		
		\path (fr1.north) +(6.25,1.75) 
		node (asrs) {\textbf{Block}};
		
		\path (fr1.north) +(9,2.15) 
		node (asrs) {\textbf{Distorted Sequences}};
		
		\path (fr1.north) +(9,1.75) 
		node (asrs) {\textbf{(Original Resolution)}};
		
		\end{tikzpicture}
    }
\end{minipage}
\centering\caption{The workflow used to obtain synthetically distorted sequences.}
\label{fig:syn_generate}
\end{figure*}

Blind assessment of perceptual video quality is challenging due to the extreme diversity of spatio-temporal artifacts that may be present. Moreover, the perception of these distortions is deeply affected by the video content, adding additional difficulty to the VQA problem. Although NR-IQA models can be used to conduct VQA in a simple way by computing quality scores at the frame-level, then temporally pooling them, this often neglects vital temporal quality information. Prior authors \cite{seshadrinathan2007structural,tu2020comparative} have demonstrated the significance of change and motion in VQA. Thus, to obtain better performance, VQA models need to account for the spatio-temporal nature of video distortions and content. Early VQA models used extensive domain knowledge about video artifacts in their feature extraction frameworks, employing regressors to map features to quality scores. Using features derived from Natural Scene Statistics (NSS) and distorted image and video statistics models \cite{moorthy2010statistics} has been a popular example employing this approach, whereby the videos are first transformed to a band-pass domain, where deviations from expected statistical regularities of the band-pass coefficients due to distortion are used to capture perceptual aspects of quality. Exemplar VQA models include VBLIINDS \cite{saad2014blind}, Li \etal \cite{li2016spatiotemporal}, which uses bandpass DCT coefficients, and BRISQUE \cite{mittal2012no}, NIQE \cite{mittal2013making}, VIIDEO \cite{mittal2015completely} and Dendi \etal \cite{dendi2020no} which employ mean subtracted contrast normalized (MSCN) bandpass coefficients to obtain perceptually relevant quality features. V-CORNIA \cite{xu2014no} employs codebook based unsupervised feature learning to conduct frame quality prediction, and hysteresis pooling \cite{seshadrinathan2011temporal} to obtain final video quality predictions. Korhonen proposed a two-level quality model TLVQM \cite{korhonen2019two} containing a comprehensive set of low complexity features extracted from all frames present in the video, along with high complexity features such as spatial activity, exposure etc. computed on a representative subset of the frames. VIDEVAL \cite{tu2021ugc} uses an ensemble approach whereby features from multiple NR models were carefully selected, yielding a feature set that was shown to perform well on UGC videos.

Recently, end-to-end trained Deep Neural Networks (DNN) have achieved significant performance gains on NR-VQA problems. Since there is lack of large scale VQA datasets to pretrain DNNs from scratch, most of the DNN based models employ transfer learning, whereby a pretrained DNN (pretrained on large datasets like ImageNet \cite{russakovsky2015imagenet} or Kinetics-400 \cite{kay2017kinetics}) is fine-tuned in a supervised manner on human opinion scores. The RAPIQUE model proposed by Tu \etal \cite{tu2021rapique} employs a hybrid set of features derived from NSS and an ImageNet pretrained CNN. Zhang \etal \cite{zhang2018blind} used FR-VQA model predictions as weak labels to train a Convolution Neural Network (CNN), then used this weakly trained CNN to conduct NR-VQA. V-MEON \cite{liu2018end} uses a multi-task training objective, where codec classification and video quality regression are jointly optimized. VSFA \cite{li2019quality} employs ImageNet pretrained CNN to extract content dependent features, along with a Gated Recurrent Unit (GRU) to model temporal memory effects. Ying \etal \cite{ying2021patch} demonstrated that features expressive of both combining local patch quality and global video quality can yield significant performance gains. GSTVQA \cite{chen2021learning} employs a pyramid temporal aggregation of short-term and long-term memory effects to achieve efficient quality prediction. Li \etal \cite{li2022blindly} employs transfer learning on IQA databases, along with a list-wise ranking loss objective to achieve competitive performance. All of the above methods employ supervised fine-tuning techniques to obtain good performances. By contrast, we focus on \textit{self-supervised} feature learning with no requirement of human quality labels during training, and no additional fine-tuning during evaluation.

\subsection{Self-Supervised Learning}
Self-supervised techniques rely on generating supervisory signals from unlabeled data. This is achieved by using an auxiliary/proxy task from which labels can be easily obtained. The auxiliary task is usually closely related to the original task, so that model trained for the auxiliary problem can be easily applied to the original task. Example tasks that have been previously studied include rotation prediction \cite{gidaris2018unsupervised}, predicting the relative positions of image patches \cite{doersch2015unsupervised}, image colorization \cite{zhang2016colorful,larsson2017colorization}, and image inpainting \cite{pathak2016context}. Recent works employing contrastive learning \cite{dosovitskiy2014discriminative,oord2018representation,bachman2019learning,chen2020simple,he2020momentum} have used instance discrimination to distinguish a data samples and augmented versions of them from other input data samples, to achieve significant performance gains on multiple computer vision problems.

Several of the above ideas have been extended to learn video representations. Fernando \etal \cite{fernando2016rank} employed rankings of frame level video features in chronological order as a proxy, to learn representations for action recognition. In \cite{lee2017unsupervised}, a temporal coherency measure is used as a supervisory signal to learn video representations. Han \etal \cite{han2019video} proposed Dense Predictive Coding (DPC) framework by recurrently predicting future representations. In \cite{yao2020video}, video playback rate perception is used as an auxiliary task for learning video embeddings. The successes of contrastive learning on pictures have been extended to videos in \cite{tao2020self,qian2021spatiotemporal,pan2021videomoco,feichtenhofer2021large}. A recently proposed self-supervised VQA model called CSPT \cite{chen2021contrastive} employs distortion and content specific contrastive loss to predict features of future video frames. Further, a cross-entropy loss is added to discriminate different distortions. Our proposed method greatly differs from CSPT, since we employ a pure contrastive loss trained model with the main goal of discriminating different distortion types and perceptual degrees of degradation. This formulation achieves significantly better performance than CSPT.

\section{Method}
\label{sec:Method}

Our goal here is to learn generic video representations that may be used to predict video quality, with no requirement of having ground-truth quality labels during training. Given a video $x$ of dimensions $H \times W \times T$, where H, W and T correspond to height, width and number of frames respectively, we are interested in learning a transformation $f:\mathbb{R}^{3 \times H \times W \times T} \mapsto \mathbb{R}^d$ that maps $x$ to a $d-$dimensional representation $h$. Prior NSS inspired VQA models have employed band-pass transformations such as the DCT \cite{saad2014blind,li2016spatiotemporal}, local mean-subtraction \cite{mittal2012no,mittal2015completely,dendi2020no} and so on, to obtain perceptually relevant quality features. Recent models employing deep learning based transforms \cite{li2019quality,ying2021patch,chen2021contrastive,li2022blindly} have also been highly successful in capturing perceptual video distortions.

Our proposed training workflow is illustrated in Fig. \ref{fig:overview_video}. The pipeline can be divided into two parts: (a) Spatial component - spatial features from CONTRIQUE \cite{madhusudana2021image} model are extracted from every video frame. (b) Temporal component - a recurrent unit maps the observed temporal variations of spatial features to a $d-$dimensional feature vector $h$. In the following sections, each component present in the pipeline is described in detail.

\subsection{Proxy/Auxiliary Task}
\label{sec:aux_task}
In self-supervised learning the original problem is modified into an alternate but closely related auxiliary task, for which ground-truth labels are easily available or can be obtained. A model is then trained to solve the auxiliary task, then during the inference stage the trained model is evaluated on the original problem. Inspired by the success of CONTRIQUE in capturing still picture distortions, we follow the idea of learning representations that can discriminate different distortion types, as well as the strength of degradations. Thus, the auxiliary task is a classification problem, where each class contains videos corresponding to similar distortion type and degree of degradation.

\subsubsection*{\textbf{Generating data for auxiliary task}} Since the auxiliary task requires a database of videos spanning different distortion types and degradation levels, the first step is to create a database of synthetically distorted videos. In this work we mainly focus on three widely observed synthetic video distortions: compression, scaling, and temporal artifacts arising due to changes in frame rate. To obtain holistic learning of different artifact types, the created dataset should contain videos which reflect individual as well as the combined effects of the above distortion types. A pristine high quality source sequence $x$ is provided as input to the distorted video generator module. A high level block diagram illustrating the workflow used to obtain synthetically distorted sequences is shown in Fig. \ref{fig:syn_generate}. The description of each of the blocks shown in Fig. \ref{fig:syn_generate} is provided below.

\begin{itemize}
    \item \textbf{Frame Rate Control}: This block modifies the frame rate of the source sequence, and generates videos having frame rates equal to or smaller than that of the source video. Similar to the LIVE-YT-HFR database \cite{madhusudana2021liveythfr}, the frame rates are chosen from the set $F = \{24, 30, 60, 82, 98, 120 \}$. The values in the set $F$ denote frames per second (fps). Note that we only choose values from $F$ that are either equal to or smaller than the source frame rate. For example, for a source video having 120 fps, all of the values in $F$ are chosen, while for a 60 fps source video only three values (60, 30 and 24 fps) are used. The low-frame rate versions were obtained by dropping frames, method by modifying the \textit{fps} filter present in FFmpeg \cite{ffmpeg}.
    \item \textbf{Scaling}: The output videos from the frame rate control block are fed as input to the scaling module, which modifies the spatial resolutions of the videos. Scaling block consists of a downsampling module preceding a compression encoder, both followed by an upsampler, so that the resulting video has the same resolution as the original. This is done to replicate scenarios usually seen in streaming and social media platforms, where videos are encoded after reducing them to lower spatial resolutions before transmission. Encoding at lower resolutions might be preferable when either the content has low complexity leading to minimal loss when downsampled, or when there are limitations on the bandwidth available for transmission \cite{knoche2005can,cermak2011relationship,georgis2015reduced}. The Lanczos kernel was used when downsampling and upsampling the videos. To ensure that the aspect ratios of the videos remained unchanged after downsampling, the resizing factors were chosen from the set $D = \{1, 2, 4, 8\}$. We also did not include downsampling factors that would have resulted in videos having resolutions less than 240p ($320 \times 240$), since these are rarely used. For example, a 1080p source content will therefore only have three scaled versions: $1980 \times 1080$, $960 \times 540$ and $480 \times 270$.
    \item \textbf{Compression}: Each of the subsampled sequences were subjected to 5 levels of VP9 compression using single pass Ffmpeg encoding in constant quality mode. Constant Rate Factor (CRF) values were varied to obtain different compression levels. The CRF values were chosen from the set $CF = \{ \text{lossless}, 24, 36, 48, 63 \}$, and the same CRF values were applied to all of the source contents. The values in the set $CF$ were chosen to include extreme cases (lossless and CRF=63, the latter corresponding to the highest possible compression level in VP9) as well as the wide range of compression levels between. Here, we make the simplifying assumption that the CRF values can be used as proxies to measure the degree of compression.
\end{itemize}

Frame rate changes may introduce temporal artifacts such as judder and strobing, while spatial downsampling and upsampling procedures can introduce scaling distortions. Our approach to generating distortions, coupled with compression, aims to realize realistic video representation currently available over internet, social media and streaming platforms. These kind of videos are included so that the learned representations will be sensitive to the synthesized distortion types.

Let a pristine source sequence $x$ be subjected to the distortion pipeline shown in Fig. \ref{fig:syn_generate}, resulting in a distorted sequence $x^{fdc}$ where $f \in F, d \in D$, and $c \in CF$. We define the auxiliary task of identifying the values $f$, $d$ and $c$ on a given distorted video $x^{fdc}$. This defines a classification problem having $|F| \times |D| \times |CF|$ classes, where $|.|$ is the set cardinality. In our experimental setting, $|F| = 6, |D| = 4$, and $|CF| = 5$, resulting in a total of 120 classes. 

\begin{table*}[t]
\caption{Summary of VQA database characteristics used for evaluation.}
\centering
\label{table:dataset_characteristics}
    \begin{tabular}{|c||c|c|c|c|c|}
        \hline
        Database & Number of Videos & Number of Scenes & Resolution & Time Duration & Distortion Type \\ \hline \hline
        KoNViD \cite{hosu2017konstanz} & 1200 & 1200 & 540p & 8s & ~ \\ 
        LIVE-VQC \cite{sinno2018large} & 585 & 585 & 240p-1080p & 10s & ~ \\ 
        YouTube-UGC \cite{wang2019youtube} & 1380 & 1380 & 360p-4K & 20s & Diverse distortions \\ 
        LSVQ \cite{ying2021patch} & 39075 & 39075 & 99p-4K & 5-12s & (authentic / UGC) \\ 
        CVD2014 \cite{nuutinen2016cvd2014} & 234 & 5 & 480p & 10-25s & ~ \\ 
        LIVE-Qualcomm \cite{ghadiyaram2017capture} & 208 & 54 & 1080p & 15s & ~ \\ \hline
        LIVE-VQA \cite{seshadrinathan2010study} & 160 & 10 & $768 \times 432$ & 10s & ~ \\ 
        CSIQ-VQA \cite{vu2014vis3} & 228 & 12 & $832 \times 480$ & 10s & Compression, \\ 
        ETRI-LIVE STSVQ \cite{lee2021subjective} & 437 & 15 & 540p-4K & 4-7s & scaling, transmission \\ 
        AVT-VQDB-UHD-1 \cite{rao2019avt} & 432 & 10 & 360p-4K & 8-10s & ~ \\ \hline
        LIVE-YT-HFR \cite{madhusudana2021liveythfr} & 480 & 16 & 1080p-4K & 8-10s & Frame rate, \\ 
        BVI-HFR \cite{mackin2018study} & 88 & 22 & 1080p & 10s & compression \\
        \hline
    \end{tabular}
\end{table*}

\subsection{Contrastive Loss}
Motivated by the success of using contrastive loss for learning representations in CONTRIQUE \cite{madhusudana2021image}, we follow a similar approach to capture space-time video distortions in CONVIQT. To obtain embeddings, we defined a deep model consisting of three components: a spatial encoder, a recurrent unit and a projector. The spatial encoder extracts features on every frame of the input video, then these features are fed to the recurrent unit. The recurrent unit extracts temporal information from these features. The projector is a fully connected network (FCN) that reduces the dimensionality of the features produced by the recurrent unit. Since CONTRIQUE representations have been shown to achieve impressive performance in capturing spatial artifacts, we employ CONTRIQUE with frozen weights as a spatial encoder, and the CONTRIQUE weights are not modified during training. Thus, the recurrent unit and the projector network are the trainable components of CONVIQT. Let $f(.)$ and $g(.)$ denote the recurrent unit and the projector network, respectively. For a given video $x \in \mathbb{R}^{3 \times H \times W \times T}$

\begin{align}
    y^{(t)} &= CONTRIQUE(x^{(t)}), \qquad t = \{1,2,\hdots,T\} \\
    h^{(t)} &= f(y^{(t)},h^{(t-1)}), \quad h = \frac{1}{T} \sum_{t=1} ^T h^{(t)}, \label{eqn:h_pool}\\
    z &= g(h) \quad h \in \mathbb{R}^d, z \in \mathbb{R}^K
\end{align}
where $x^{(t)}$ is the $t^{th}$ frame of video $x$, and $h$ is the $d$-dimensional temporal pooled output from the recurrent unit. The representation $h$ is $L_2$ normalized \cite{chen2020simple,he2020momentum,madhusudana2021image} before being fed to the projector. The projector reduces the feature dimensions from $d$ to $K$ ($K$ is a hyperparameter in this setting). The objective is to obtain similar embeddings $z$ for videos belonging to the same class. The cosine distance between the representations $z_i$ and $z_j$, given by $\phi(z_i,z_j) = z_i^T z_j/||z_i||_2 ||z_j||_2$ is used as a similarity measure. The loss function is similar to the supervised contrastive loss proposed in \cite{khosla2020supervised}. For a video $x_i$ it is defined as 
\begin{align}
    \mathcal{L}_i ^{syn} = \frac{1}{|V(i)|} \sum_{j \in P(i)} -\log \frac{\exp(\phi(z_i,z_j)/\tau)}{\sum_{m=1}^N \mathbbm{1}_{m \neq i}\exp(\phi(z_i,z_m)/\tau)},
    \label{eqn:cont_loss_sup}
\end{align}
where $N$ is the number of videos present in the batch, $\mathbbm{1}$ is the set indicator function, $\tau$ is a temperature parameter, $V(i)$ is a set containing video indices belonging to the same class as $x_i$ (but excluding the index $i$), and $|V(i)|$ is its cardinality. For example, if $x_i$ is a video which was encoded at 540p, 60fps and CRF=24, $V(i)$ will contain indices of all videos (excluding index $i$) which were encoded at 540p, 60fps and CRF=24 setting. There exists transformations of a video $x_i$ (video transformations are discussed in Sec. \ref{sec:multi_scale_sec}) guaranteeing that $V(i)$ contains at least one sample. Note that a distorted video's class is uniquely determined by its frame rate, spatial subsampling factor, and compression level, and is independent of the native resolution of the source video. For example, a 4K 120fps source encoded at 1080p, 60 fps, and CRF=48 will belong to the same class as that of a 1080p 120 fps source encoded at 540p, 60 fps, and CRF=48, since both are spatially downsampled twice before encoding. This resolution independent class setting enables us to employ source videos spanning multiple spatial resolutions during training, without requiring any need to resize videos to predetermined sizes. The objective defined in (\ref{eqn:cont_loss_sup}) is a normalized temperature-scaled cross entropy (NT-Xent), and is calculated between every pair of videos present in the batch. 

\subsection{Multiscale Learning and Temporal Transformations}
\label{sec:multi_scale_sec}
Images, and by extension videos are inherently multi-scale, and perceived quality is influenced by local details as well as by global characteristics. Prior VQA models \cite{mittal2012no,saad2014blind,tu2021rapique} have achieved significant gains in performance by employing multi-scale representations. In CONVIQT, we use two scales: native resolutions, and half-scale resolutions obtained by downscaling, twice along both spatial dimensions. An anti-aliasing filter is used before downscaling as shown in Fig. \ref{fig:overview_video} to avoid aliasing artifacts. Note that this downscaling operation preserves the aspect ratio of the input video, which is desirable since changing this could modify the perceived quality of the input video. 

The aim of the objective function in (\ref{eqn:cont_loss_sup}) is to obtain discriminative representations that can differentiate different video classes, and also demonstrate invariance to quality preserving transformations. Operations on videos which preserve video quality we will collectively refer to as quality preserving transforms. Depending on the coordinates on which video transforms are applied, they can be further categorized as spatial and temporal transforms. Spatial transforms are applied independently on each video frame, while temporal transforms are applied along the time axis. Since the CONTRIQUE model is only used to obtain spatial features, and during training of CONTRIQUE, spatial transforms such as horizontal flipping, color-space conversion were employed, we do not include any spatial transforms when training CONVIQT. 

We employ temporal band-pass transforms inspired by similar successes \cite{madhusudana2021st,madhusudana2021high,madhusudana2022making,tu2021rapique,zheng2022faver} capturing temporal distortions. Temporal band-pass coefficients demonstrate reliable statistical regularities on very high quality videos, but these are predictably disturbed by the presence of distortions. We use them to access temporal quality information over multiple scales (subbands). Similar to \cite{madhusudana2021st}, we employ three temporal band-pass filters: Haar, Daubechies-2 (db2), and Biorthogonal-2.2 (bior2.2) with wavelet packet decomposition \cite{coifman1992entropy}. In each case, we use three levels of wavelet decomposition resulting in 7 subbands (ignoring the low pass response). During training, the wavelet filter and subbands were randomly chosen for each input video. The transformed videos at two spatial scales were then fed as input to the CONTRIQUE model to obtain spatial representations. Since the CONTRIQUE weights remain unchanged during CONVIQT training, the spatial representations of all the transformed videos present in the training data can be pre-computed and stored in the disk. This significantly reduces the overall training time of CONVIQT.

To facilitate batch training, the videos are randomly cropped along the temporal axis to length of $T$ frames. This ensures that every video in the batch contained the same number of frames, which is necessary when training deep networks. We presume that this cropping operation did not alter the distortion class, and that the cropped portion inherits the class of the original video. Temporal cropping is performed on every video at both full-scale and at half-scale. We ensured that all of the videos in the training set contain at least $T$ frames to avoid zero padding.
\looseness=-1

\subsection{Realistic Distortions}
In the contrastive objective (\ref{eqn:cont_loss_sup}), prior knowledge of the distortion types and degradation degrees were used to learn video representations. However, this kind of information is usually not readily available for UGC videos, despite the massive volumes of streamed and shared on social media platforms like YouTube, Facebook, Instagram and TikTok. Given their consumer ubiquity, it is vital that the learned embeddings be sensitive to the realistic artifacts found on UGC sequences, which often include mixtures of multiple unknown distortion types. Thus, the classes used in (\ref{eqn:cont_loss_sup}) are only poorly applicable to UGC videos. Since identifying the distortion type and degree of degradation of UGC videos is not easy task (even for human eyes), we instead simplify the problem by treating each UGC video as a unique class containing a distinctive blend of multiple distortions, separate and different from other UGC videos, as well as from synthetically distorted sequences. Thus, given a UGC video $x_i$, only its scaled (and temporally transformed) version $x_j$ belongs to the same class. The loss objective in (\ref{eqn:cont_loss_sup}) is modified to reflect this change as
\begin{align}
    \mathcal{L}_i ^{UGC} =  -\log \frac{\exp(\phi(z_i,z_j)/\tau)}{\sum_{m=1}^N \mathbbm{1}_{m \neq i}\exp(\phi(z_i,z_m)/\tau)}.
    \label{eqn:cont_loss_ugc}
\end{align}
This is essentially the instance discrimination objective used in prior work \cite{chen2020simple,he2020momentum}. Combining (\ref{eqn:cont_loss_sup}) and (\ref{eqn:cont_loss_ugc}), the overall training objective is then
\begin{align}
    \mathcal{L} = \frac{1}{N} \sum_{i = 1} ^N \mathbbm{1}_{(x_i \notin UGC)} \mathcal{L}_i ^{syn} + \mathbbm{1}_{(x_i \in UGC)}\mathcal{L}_i ^{UGC},
    \label{eqn:cont_loss_total}
\end{align}
where $N$ is the total number of videos present in the batch, and $\mathbbm{1}$ is the indicator function determining whether the input video is UGC. To avoid any bias during training, at each iteration we sampled equal numbers of synthetic and authentically distorted videos.
\looseness=-1

\begin{table*}[t]
\caption{Performance comparison of CONVIQT against different NR models on VQA databases containing \textbf{authentic} distortions. In each column, the three best models are boldfaced. Entries marked '-' denote that the results are not available.}
\label{table:authentic_VQA}
\centering
 \footnotesize
 \resizebox{1\textwidth}{!}{
    \begin{tabular}{|c||c||cc|cc|cc|cc|cc|}
        \hline
        \multirow{3}{*}{Method} & \multirow{3}{*}{Model Type} & \multicolumn{2}{|c|}{\multirow{2}{*}{KoNViD \cite{hosu2017konstanz}}}  & \multicolumn{2}{|c|}{\multirow{2}{*}{LIVE-VQC \cite{sinno2018large}}} & \multicolumn{2}{|c|}{\multirow{2}{*}{YouTube-UGC \cite{wang2019youtube}}} & \multicolumn{4}{|c|}{LSVQ \cite{ying2021patch}}\\ \cline{9-12}
        ~ & ~ & ~ & ~ & ~ & ~ & ~ & ~ & \multicolumn{2}{|c|}{Test} & \multicolumn{2}{|c|}{Test-1080p} \\ \cline{3-12}
        ~ & ~ & SROCC$\uparrow$ & PLCC$\uparrow$ & SROCC$\uparrow$ & PLCC$\uparrow$ & SROCC$\uparrow$ & PLCC$\uparrow$ & SROCC$\uparrow$ & PLCC$\uparrow$ & SROCC$\uparrow$ & PLCC$\uparrow$ \\ \hline \hline
        BRISQUE \cite{mittal2012no} & \multirow{5}{*}{\shortstack[c]{Traditional/ \\ Handcrafted \\ Features}} & 0.649 & 0.651 & 0.593 & 0.624 & 0.393 & 0.407 & 0.579 & 0.576 & 0.497 & 0.531 \\
        VBLIINDS \cite{saad2014blind} & ~ & 0.706 & 0.701 & 0.681 & 0.699 & 0.534 & 0.540 & - & - & - & - \\
        TLVQM \cite{korhonen2019two} & ~ & 0.758 & 0.759 & 0.787 & 0.794 & 0.656 & 0.647 & 0.772 & 0.774 & 0.589 & 0.616 \\
        VIDEVAL \cite{tu2021ugc} & ~ & 0.770 & 0.770 & 0.743 & 0.747 & 0.776 & 0.771 & 0.794 & 0.783 & 0.545 & 0.554 \\ 
        RAPIQUE \cite{tu2021ugc} & ~ & 0.788 & 0.805 & 0.741 & 0.761 & 0.747 & 0.756 & - & - & - & - \\ \hline
        VSFA \cite{li2019quality} & \multirow{5}{*}{\shortstack[c]{Supervised \\ Pretraining}} & 0.794 & 0.798 & 0.717 & 0.770 & 0.787 & 0.788 & 0.801 & 0.796 & 0.675 & 0.704 \\
        PVQ \cite{ying2021patch} & ~ & 0.791 & 0.786 & \textbf{0.827} & \textbf{0.837} & - & - & 0.814 & 0.816 & \textbf{0.686} & \textbf{0.708} \\
        GSTVQA \cite{chen2021learning} & ~ & 0.814 & 0.825 & 0.784 & 0.787 & - & - & - & - & - & - \\
        Li \etal \cite{li2022blindly} & ~ & \textbf{0.835} & \textbf{0.833} & \textbf{0.841} & \textbf{0.839} & \textbf{0.825} & \textbf{0.817} & \textbf{0.852} & \textbf{0.853} & \textbf{0.771} & \textbf{0.787} \\ 
        Resnet-50 \cite{he2016deep} & ~ & 0.820 & 0.823 & 0.770 & 0.804 & 0.805 & 0.811 & 0.820 & 0.817 & \textbf{0.706} & \textbf{0.735} \\ \hline
        CSPT \cite{chen2021contrastive} & \multirow{4}{*}{\shortstack[c]{Unsupervised \\ Pretraining}} & 0.701 & 0.673 & 0.622 & 0.630 & - & - & - & - & - & - \\
        CONTRIQUE \cite{madhusudana2021image} & ~ & \textbf{0.844} & \textbf{0.842} & \textbf{0.815} & \textbf{0.822} & \textbf{0.825} & \textbf{0.813} & \textbf{0.828} & \textbf{0.826} & 0.662 & 0.697 \\
        CONVIQT & ~ & \textbf{0.851} & \textbf{0.849} & 0.808 & 0.817 & \textbf{0.832} & \textbf{0.822} & \textbf{0.821} & \textbf{0.820} & 0.668 & 0.702 \\
        
        \hline
    \end{tabular}
    }
\end{table*}

\subsection{Evaluating Representations}
The learned representations are evaluated with respect to their video quality prediction power by correlating their responses against human opinion scores. After the CONVIQT model is trained, the projector network $g(.)$ is discarded and the outputs of the recurrent unit $h = f(y^{(1)},\hdots,y^{(t)})$ are used as video quality embeddings. A regularized linear (ridge) regressor is trained on top of the frozen CONVIQT model in order to map embeddings to video quality scores. The regression expression is given by
\begin{align*}
    y^{(t)}_j &= CONTRIQUE(x^{(t)}_j), \hfill t = \{1,\hdots,T\},  j = \{1,\hdots,N\} \\
    h^{(t)}_j &= f(y^{(t)}_j,h^{(t-1)}_j), \quad h_j = \frac{1}{T} \sum_{t=1} ^T h^{(t)}_j, \\
    p &= W h, \quad W^* = \argmin_{W} \sum_{j=1} ^N (GT_j - p_j)^2 + \lambda \sum_{i=1} ^d W_i^2,
\end{align*}
where $GT_j$ are ground-truth quality scores, $p_j$ are predicted scores, $W$ is a trainable vector having the same dimensions as $h$, $\lambda$ is a regularization parameter, $d$ is the number of dimensions of $h$, and $N$ is the number of videos present in the VQA database training set. Similar to training, multiscale processing is conducted during evaluation, and features are computed at two scales: full-scale and half-scale, with the final embedding being a concatenation of embeddings from both scales. During the evaluation phase, all of the features were calculated without performing any additional temporal transforms or cropping. Note that when the ridge regressor was trained, the weights of the recurrent unit (as well as those of CONTRIQUE) were frozen without any additional fine-tuning. Although fine-tuning could lead to better correlations, it modifies the encoder weights, and would not be a true measure of the effectiveness of the unsupervised learning process. However, we show in Sec. \ref{sec:correlation} that even without fine-tuning, CONVIQT obtains competitive performance against SOTA VQA models.

\begin{table}[t]
\caption{Performance comparison of CONVIQT against different NR models on the CVD2014 and LIVE-Qualcomm databases. In each column, the three best models are boldfaced. Entries marked '-' denote that the results are not available.}
\label{table:authentic_VQA_qual}
\centering
    \begin{tabular}{|c||c|c|c|c|}
        \hline
        \multirow{2}{*}{Method} & \multicolumn{2}{|c|}{CVD2014 \cite{nuutinen2016cvd2014}} & \multicolumn{2}{|c|}{LIVE-Qualcomm \cite{ghadiyaram2017capture}} \\ \cline{2-5}
        ~ & SROCC$\uparrow$ & PLCC$\uparrow$ & SROCC$\uparrow$ & PLCC$\uparrow$ \\ \hline \hline
        BRISQUE \cite{mittal2012no} & 0.790 & 0.804 & 0.552 & 0.598 \\ 
        VBLIINDS \cite{saad2014blind} & 0.795 & 0.806 & 0.570 & 0.626 \\ 
        TLVQM \cite{korhonen2019two} & 0.779 & 0.790 & 0.785 & 0.815 \\ 
        VIDEVAL \cite{tu2021ugc} & 0.814 & 0.832 & 0.670 & 0.705 \\ 
        RAPIQUE \cite{tu2021rapique} & 0.807 & 0.823 & 0.665 & 0.691 \\ \hline
        VSFA \cite{li2019quality} & 0.\textbf{850} & \textbf{0.869} & 0.708 & 0.774 \\
        GSTVQA \cite{chen2021learning} & 0.831 & 0.844 & \textbf{0.801} & \textbf{0.825} \\
        Li \etal \cite{li2022blindly} & \textbf{0.862} & \textbf{0.882} & \textbf{0.833} & \textbf{0.837} \\
        Resnet-50 \cite{he2016deep} & 0.835 & 0.838 & 0.650 & 0.691 \\ \hline
        CSPT \cite{chen2021contrastive} & 0.653 & 0.691 & - & - \\
        CONTRIQUE \cite{madhusudana2021image} & 0.826 & 0.836 & 0.765 & 0.777 \\
        CONVIQT & \textbf{0.858} & \textbf{0.837} & \textbf{0.797} & \textbf{0.802} \\
        \hline
    \end{tabular}
\end{table}

\section{Experiments and Results}
\label{sec:experiments}
In this section we evaluate the performance of CONVIQT by carrying out a series of experiments. First we will describe the experiment settings used when pretraining CONVIQT, the evaluation protocol, and the VQA models used for comparison. We also report the performance of CONVIQT against SOTA VQA models on multiple VQA datasets. To investigate the generalizability of the CONVIQT representations, we further performed cross-dataset evaluations. We also performed a collection of ablation experiments to analyze the significance of distortion types present in the pretraining data, and the impact of the various temporal transforms. 

\begin{table*}[t]
\caption{Performance comparison of CONVIQT against different NR models on VQA databases containing \textbf{synthetic} distortions. In each column, the three best models are boldfaced. Entries marked '-' denote that the results are not available.}
\label{table:synthetic_VQA}
    \centering
    \footnotesize
    \begin{tabular}{|c||c||cc|cc|cc|cc|}
        \hline
        \multirow{2}{*}{Method} & \multirow{2}{*}{Model Type} & \multicolumn{2}{|c|}{LIVE-VQA \cite{seshadrinathan2010study}} & \multicolumn{2}{|c|}{CSIQ-VQA \cite{vu2014vis3}} & \multicolumn{2}{|c|}{ETRI-LIVE STSVQ \cite{lee2021subjective}} & \multicolumn{2}{|c|}{AVT-VQDB-UHD-1 \cite{rao2019avt}} \\
        \cline{3-10}
        ~ & ~ & SROCC$\uparrow$ & PLCC$\uparrow$ & SROCC$\uparrow$ & PLCC$\uparrow$ & SROCC$\uparrow$ & PLCC$\uparrow$ & SROCC$\uparrow$ & PLCC$\uparrow$\\ \hline \hline
        BRISQUE \cite{mittal2012no} & \multirow{4}{*}{\shortstack[c]{Traditional/ \\ Handcrafted \\ Features}} & 0.567 & 0.571 & 0.654 & 0.672 & 0.351 & 0.341 & 0.838 & 0.815 \\ 
        VBLIINDS \cite{saad2014blind} & ~ & \textbf{0.724} & \textbf{0.748} & 0.784 & 0.771 & 0.480 & 0.534 & \textbf{0.875} & \textbf{0.876} \\ 
        TLVQM \cite{korhonen2019two} & ~ & \textbf{0.733} & \textbf{0.751} & \textbf{0.794} & \textbf{0.774} & 0.426 & 0.423 & 0.832 & 0.823 \\ 
        VIDEVAL \cite{tu2021ugc} & ~ & 0.475 & 0.463 & 0.368 & 0.300 & 0.360 & 0.332 & 0.664 & 0.609 \\ 
        VSFA \cite{li2019quality} & \multirow{2}{*}{\shortstack[c]{Supervised \\ Pretraining}} & 0.700 & 0.727 & \textbf{0.798} & \textbf{0.781} & - & - & - & -\\
        Resnet-50 \cite{he2016deep} & ~ & 0.442 & 0.385 & 0.703 & 0.686 & \textbf{0.889} & \textbf{0.902} & {0.865} & {0.880}  \\ \hline
        CSPT \cite{chen2021contrastive} & \multirow{3}{*}{\shortstack[c]{Unsupervised \\ Pretraining}} & \textbf{0.713} & \textbf{0.726} & 0.751 & 0.749 & - & - & - & - \\
        CONTRIQUE \cite{madhusudana2021image} & ~ & 0.636 & 0.623 & \textbf{0.795} & \textbf{0.775} & \textbf{0.931} & \textbf{0.930} & \textbf{0.873} & \textbf{0.920}  \\
        CONVIQT & ~ & 0.622 & 0.595 & 0.766 & 0.749 & \textbf{0.939} & \textbf{0.936} & \textbf{0.867} & \textbf{0.906} \\
        \hline
    \end{tabular}
\end{table*}

\subsection{Experimental Settings}
\subsubsection*{\textbf{Pretraining Data}}
The pretraining data contains a mix of videos afflicted with synthetic and realistic artifacts.

\begin{itemize}
    \item Synthetic distortions: To learn synthetic artifacts, we first generated data as detailed in Sec. \ref{sec:aux_task}. The high quality pristine contents required to generate the distorted sequences were obtained from 5 different sources as described below.
    \begin{enumerate}
        \item Waterloo GRD database \cite{duanmu2020characterizing}: Contains 1000 videos of 1080p resolution captured at 24/30 fps and 10s in duration. We randomly sampled 500 source sequences from this database.
        \item Dareful \cite{dareful}: The Dareful website contains open source high quality 4K stock video footages shot at 30 fps. We used 46 sequences from this platform.
        \item REDS \cite{nah2019ntire}: The REDS dataset contains videos of 720p resolution captured at 120 fps. This dataset was originally created to benchmark video deblurring and super-resolution algorithms. We used a total of 270 (240 from training set and 30 from validation set) videos from this database.
        \item MCML \cite{cheon2017subjective}: MCML is a VQA dataset containing 240 distorted videos obtained from 10 reference sequences of 4K resolution captured at 30 fps. We used all of the reference sequences from this dataset.
        \item UVG \cite{mercat2020uvg}: Ultra Video Group (UVG) dataset contains 16 diverse 4K contents at 50/120 fps. To avoid overlap of contents between the VQA databases used for evaluation, we used only three contents of 4K resolution at 120 fps: 'Bosphorus', 'Lips' and 'ShakeNDry'.
    \end{enumerate}
    Thus, 829 pristine sequences spanning multiple resolutions and frame rates were used to generate a total of 30855 synthetically distorted sequences.
    \item Authentic distortions: To capture realistic distortions we used videos from the training set of Kinetics-400 dataset \cite{kay2017kinetics}. Kinetics-400 is a human action video dataset containing 400 human action classes where each video is of duration around 10s. The training set of Kinetics-400 contains 240K videos of which we randomly sampled 30K videos, discarding all labels before training. 
\end{itemize}

Thus, a total of 60K videos were used to pretrain CONVIQT. Additionally, we ensured that there was no content overlap between the pretraining sequences and the sequences present in VQA datasets used for evaluation.

\subsubsection*{\textbf{Pretraining Details}}
We employed a Gated Recurrent Unit (GRU) \cite{cho2014properties} as the recurrent unit $f(.)$ for modeling temporal information, and 2 layers of MLP as the projector $g(.)$. The GRU contains a single layer with a hidden size of 1024. The output $f(.)$ of the GRU is temporal average pooled (as shown in (\ref{eqn:h_pool})) before being fed to the MLP $g(.)$. The dimension of the final output $z$ was fixed to $K = 128$. The batch size for pretraining was $N = 1024$, with equal numbers of videos randomly sampled from synthetic and authentically distorted sequences. The spatial features from the CONTRIQUE engine were extracted at the native resolution of the input videos, while temporal crops of length $T=16$ frames were fed to the GRU to learn temporal information. During training, the videos were loaded in RGB format, and the same temporal transform was applied to all 3 channels of the input video. The transformed video was normalized to lie in the range $[0,1]$ before being fed as input. The temperature parameter present in expressions (\ref{eqn:cont_loss_sup}) and (\ref{eqn:cont_loss_ugc}) was fixed at $\tau = 0.1$. The GRU and the projector network were trained from scratch for 10 epochs using stochastic gradient descent (SGD) at a learning rate of $1.2$ using a batch size $N = 1024$. A linear warmup over the first two epochs was applied to the learning rate, followed by a cosine decay schedule without restarts \cite{loshchilov2016sgdr}. All the experiments were done in Python using the PyTorch\footnote{\url{https://pytorch.org/}} deep learning modules. 

\subsubsection*{\textbf{Evaluation Datasets}}
We evaluated CONVIQT on 12 large VQA datasets spanning both synthetic and realistic distortions. The characteristics of the VQA datasets used for evaluation are summarized in Table \ref{table:dataset_characteristics}. When using the LSVQ \cite{ying2021patch} database, which also has human-labeled space-time video patches, we only used the full videos (and corresponding human opinion scores), leaving out the video patches.

\subsubsection*{\textbf{Compared Methods}}
We compared the performance of CONVIQT against a variety of SOTA NR-VQA models. The compared models can be classified into three categories based on the feature design methodology: (a) Traditional/hand-crafted features - BRISQUE \cite{mittal2012no}, VBLIINDS \cite{saad2014blind}, TLVQM \cite{korhonen2019two}, VIDEVAL \cite{tu2021ugc}, and RAPIQUE \cite{tu2021rapique}. These models employ support vector regressors (SVRs) to map features to quality scores. (b) Supervised deep learning based models - VSFA \cite{li2019quality}, PVQ \cite{ying2021patch}, GSTVQA \cite{chen2021learning}, and Li \etal \cite{li2022blindly}. Note that RAPIQUE is a hybrid model employing a combination of NSS and deep CNN features. (c) Self-supervised deep learning based models - CSPT \cite{chen2021contrastive} and CONTRIQUE \cite{madhusudana2021image}. To numerically compare the above models, we copied the values reported by the respective authors or in the literature. For PVQ, which is capable of inferencing on video patches, we only report the performance of the model trained on videos, since video patch quality scores are not employed in our evaluation. To compare the results obtained by the supervised and unsupervised pretraining techniques, we also included a 2D Resnet-50 \cite{he2016deep} model pretrained on Imagenet \cite{russakovsky2015imagenet}. Since BRISQUE, Resnet-50 and CONTRIQUE are IQA models, the respective features were computed on each video frame then average pooled along the temporal dimension to obtain the final video embedding. These embeddings were then mapped to quality scores using an SVR on the BRISQUE features, and a linear regressor on the Resnet-50 and CONTRIQUE features. For fair comparison of CSPT with CONVIQT, we report the CSPT performance under linear evaluation.

\subsubsection*{\textbf{Evaluation Protocol}}
Spearman's rank order correlation coefficient (SROCC) and Pearson's linear correlation coefficient (PLCC) were used as evaluation metrics to compare the VQA models. A four-parameter logistic non-linearity \cite{VQEG2000} was applied to the quality predictions before PLCC was computed.

During evaluation, the videos were processed at their native resolution, and were partitioned into non-overlapping clips of $T$ continuous frames. From each clip CONVIQT produced a single feature vector as output, and the features from all clips were average pooled to obtain the final video representations. These representations, along with the corresponding human opinion scores were used to learn the regressor weights. When learning regressor weights, each VQA dataset was randomly split into 70\%, 10\%, and 20\% partitions corresponding to training, validation, and testing sets, respectively. The validation set was used to calculate the regularization coefficient of the regressor via grid search. For VQA datasets containing multiple distorted videos belonging to the same content, the random splits were conducted in a way that ensured no content overlap between the training and testing sets. The training protocol was repeated 100 times to avoid biases towards the choice of videos present in the training set, and the median performance was reported. Due to the large size of the LSVQ dataset, the single split reported by the authors \cite{ying2021patch} was used. 

\begin{table}[t]
\caption{Performance comparison of CONVIQT against different NR models on variable frame rate VQA databases. In each column, the three best models are boldfaced.}
\label{table:HFR_VQA}
\centering
    \begin{tabular}{|c||c|c|c|c|}
        \hline
        \multirow{2}{*}{Method} & \multicolumn{2}{|c|}{LIVE-YT-HFR \cite{madhusudana2021liveythfr}} & \multicolumn{2}{|c|}{BVI-HFR \cite{mackin2018study}} \\ \cline{2-5}
        ~ & SROCC$\uparrow$ & PLCC$\uparrow$ & SROCC$\uparrow$ & PLCC$\uparrow$ \\ \hline \hline
        BRISQUE \cite{mittal2012no} & 0.319 & 0.419 & 0.260 & 0.444 \\ 
        VBLIINDS \cite{saad2014blind} & 0.391 & 0.467 & - & - \\ 
        TLVQM \cite{korhonen2019two} & 0.429 & 0.504 & 0.373 & 0.490 \\ 
        VIDEVAL \cite{tu2021ugc} & 0.474 & 0.566 & 0.344 & 0.474 \\
        RAPIQUE \cite{tu2021rapique} & 0.456 & 0.566 & 0.303 & 0.463 \\ 
        FAVER \cite{zheng2022faver} & 0.635 & 0.692 & \textbf{0.556} & \textbf{0.639} \\
        Resnet-50 \cite{he2016deep} & \textbf{0.641} & \textbf{0.704} & 0.415 & 0.411 \\ \hline
        CONTRIQUE \cite{madhusudana2021image} & \textbf{0.650} & \textbf{0.706} & \textbf{0.568} & \textbf{0.556} \\
        CONVIQT & \textbf{0.672} & \textbf{0.734} & \textbf{0.641} & \textbf{0.626} \\
        \hline
    \end{tabular}
\end{table}

\subsection{Correlation Against Human Judgments}
\label{sec:correlation}
We compared the performance of CONVIQT against SOTA VQA models on authentically distorted videos, as shown in Table \ref{table:authentic_VQA}. It may be observed from the Table that CONVIQT achieved competitive correlation values compared to the other VQA models. The compared VQA models are categorized based on the type of feature extraction techniques employed. The performance differences between CONTRIQUE and CONVIQT is indicative of the significance of the recurrent unit. This difference depends on the VQA dataset being used, but in most cases employing the recurrent unit yielded better performance. Notably, the performances of CONTRIQUE and CONVIQT were achieved without any supervised fine-tuning, unlike the other models.

\begin{table*}[t]
\caption{SROCC performance comparison to analyze different CONVIQT training protocols. \textit{syn} and \textit{UGC} indicate models trained on data containing only synthetic and authentic distortions, respectively. In each row the top performing model is boldfaced.}
\label{table:cross_data_training}
    \centering
    \footnotesize
        \begin{tabular}{|c|c||c|c|c|}
        \hline
        {Training} & {Testing} & CONVIQT-syn & CONVIQT-UGC & CONVIQT \\ \hline \hline
        YouTube-UGC & KoNViD & 0.728 & 0.750 & \textbf{0.761} \\
        KoNViD & YouTube-UGC & 0.580 & 0.601 & \textbf{0.731}\\ \hline
        ETRI-LIVE STSVQ & AVT-VQDB UHD-1 & 0.709 & 0.759 & \textbf{0.774} \\ 
        AVT-VQDB UHD-1 & ETRI-LIVE STSVQ & \textbf{0.697} & 0.607 & 0.651 \\ \hline
        LIVE-YT-HFR & AVT-VQDB UHD-1 & 0.754 & 0.730 & \textbf{0.767} \\ 
        AVT-VQDB UHD-1 & LIVE-YT-HFR & \textbf{0.593} & 0.448 & {0.556} \\ 
        \hline
    \end{tabular}
\end{table*}

Table \ref{table:authentic_VQA_qual} shows performance comparisons on the CVD2014 \cite{nuutinen2016cvd2014} and LIVE-Qualcomm \cite{ghadiyaram2017capture} datasets. These datasets contain multiple videos of the same scene impaired by authentic distortions. Thus, the train/test splits were done ensuring no content overlap between them. Again, CONVIQT obtained excellent correlations when compared to other VQA models.

The performance of CONVIQT on videos modified by synthetic distortions is analyzed in Table \ref{table:synthetic_VQA}. CONVIQT obtains superior correlations on recent VQA datasets such as ETRI-LIVE STSVQ \cite{lee2021subjective} and AVT-VQDB-UHD-1 \cite{rao2019avt}, while its performance is slightly lower on legacy datasets like LIVE-VQA \cite{seshadrinathan2010study} and CSIQ-VQA \cite{vu2014vis3}. This may be attributed to the type of distortions present. ETRI-LIVE STSVQ and AVT-VQDB-UHD-1 mainly contain scaling and compression artifacts which were well modeled in the data used to pretrain CONVIQT. The legacy datasets contain distortions such as transmission errors arising in wireless and IP networks, and packet loss errors, which were not included in the training data and could be a contributing factor in obtaining lower correlation values. 

\begin{table}[t]
\caption{Cross database SROCC comparison of VQA models. In each row the top performing model is boldfaced.}
\label{table:cross_data}
    \centering
    \footnotesize
    \resizebox{0.48\textwidth}{!}{
        \begin{tabular}{|c|c||c|c|}
        \hline
        Training & Testing & CONTRIQUE & CONVIQT \\ \hline \hline
        YouTube-UGC & KoNViD & 0.718 & \textbf{0.761} \\
        KoNViD & YouTube-UGC & 0.498 & \textbf{0.731}\\ \hline
        ETRI-LIVE STSVQ & AVT-VQDB UHD-1 & \textbf{0.785} & {0.774} \\ 
        AVT-VQDB UHD-1 & ETRI-LIVE STSVQ & {0.550} & \textbf{0.651} \\ \hline
        LIVE-YT-HFR & AVT-VQDB UHD-1 & {0.740} & \textbf{0.767} \\ 
        AVT-VQDB UHD-1 & LIVE-YT-HFR & {0.378} & \textbf{0.556} \\ 
        \hline
    \end{tabular}
    }
\end{table}

\subsection{Performance Comparison on Variable Frame Rate Videos}
Table \ref{table:HFR_VQA} compares CONVIQT against other VQA models on variable frame rate (VFR) videos. The LIVE-YT-HFR \cite{madhusudana2021liveythfr} and BVI-HFR \cite{mackin2018study} VQA databases contain videos of same content but sampled at multiple combinations of frame rates and compression factors, along with corresponding human opinion scores. VFR-VQA is a challenging problem, since quality predictors must account for subtle perceptual quality changes occuring along the temporal dimension due to frame rate changes. As comparisons we also included FAVER \cite{zheng2022faver}, which is a current SOTA no-reference VFR-VQA model. From Table \ref{table:HFR_VQA}, it may be observed that CONVIQT achieved competitive performance when compared against other NR-VQA models, indicating the frame rate discrimination capability of CONVIQT representations. We may also observe significant performance gaps between CONTRIQUE and CONVIQT on same datasets, highlighting the importance of the recurrent unit.

\subsection{Cross Dataset Evaluation}
\label{sec:cross_data}
We conducted a cross dataset analysis in Table \ref{table:cross_data}. The regressors were trained and tested on different VQA datasets. Cross data evaluations are beneficial when analyzing the generalizability of compared models. We compared CONTRIQUE and CONVIQT to study whether the inclusion of the recurrent unit yields better generalization. We chose a subset of VQA datasets from Table \ref{table:dataset_characteristics}, two containing synthetic distortions, two containing UGC videos and one containing VFR videos. From the Table, it may be observed that in most cases, CONVIQT delivered much better performance than CONTRIQUE, highlighting the importance of the recurrent unit in obtaining more generalized representations. Note that even for cross dataset evaluation, only the regressor weights were modified based on the training data, while the weights of the CONVIQT model were kept intact.

\subsection{Significance of Training Data}
We noted in Sec. \ref{sec:Method} that the pretraining data for CONVIQT contains a combination of videos afflicted by synthetic and authentic distortions. In this experiment we investigate the generalization of the learned representation when CONVIQT is pretrained with either only synthetic or only authentic artifacts. The results are reported in Table \ref{table:cross_data_training} following similar evaluation strategy as in Sec. \ref{sec:cross_data} to train and test on different VQA datasets to analyze generalizability. From the Table, it may be seen that using combined datasets yielded better performance on most train/test combinations, as compared to individual trainings indicating the complementary nature of the quality information present in the different distortion types.

\subsection{Importance of Temporal Transforms}
To obtain more robust representations, the CONVIQT pipeline contains a temporal transform module as illustrated in Fig. \ref{fig:overview_video}. In this experiment we analyzed the importance of the temporal transform by removing the module during CONVIQT training. The resulting model performance is compared in Table \ref{table:cross_data_temporal} where it may be observed that the absence of a temporal transform leads to less generalizable representations.

\begin{table}[t]
\caption{SROCC performance comparison to analyze the importance of temporal transformation. In each row top performing model is boldfaced}
\label{table:cross_data_temporal}
    \centering
    \footnotesize
    \resizebox{0.48\textwidth}{!}{
        \begin{tabular}{|c|c||c|c|}
        \hline
        \multirow{2}{*}{Training} & \multirow{2}{*}{Testing} & w/o Temporal & w/ Temporal \\ 
        ~ & ~ & Transform & Transform \\ \hline \hline
        YouTube-UGC & KoNViD & 0.739 & \textbf{0.761} \\
        KoNViD & YouTube-UGC & 0.722 & \textbf{0.731}\\ \hline
        ETRI-LIVE STSVQ & AVT-VQDB UHD-1 & 0.759 & \textbf{0.774} \\ 
        AVT-VQDB UHD-1 & ETRI-LIVE STSVQ & \textbf{0.722} & 0.651 \\ \hline
        LIVE-YT-HFR & AVT-VQDB UHD-1 & 0.716 & \textbf{0.767} \\ 
        AVT-VQDB UHD-1 & LIVE-YT-HFR & \textbf{0.599} & {0.556} \\ 
        \hline
    \end{tabular}
    }
\end{table}

\subsection{Qualitative Analysis}
\begin{figure}[t]
    \centering
    \subfloat[Predicted = 2.44, GT = 2.40]{\includegraphics[width=0.45\linewidth]{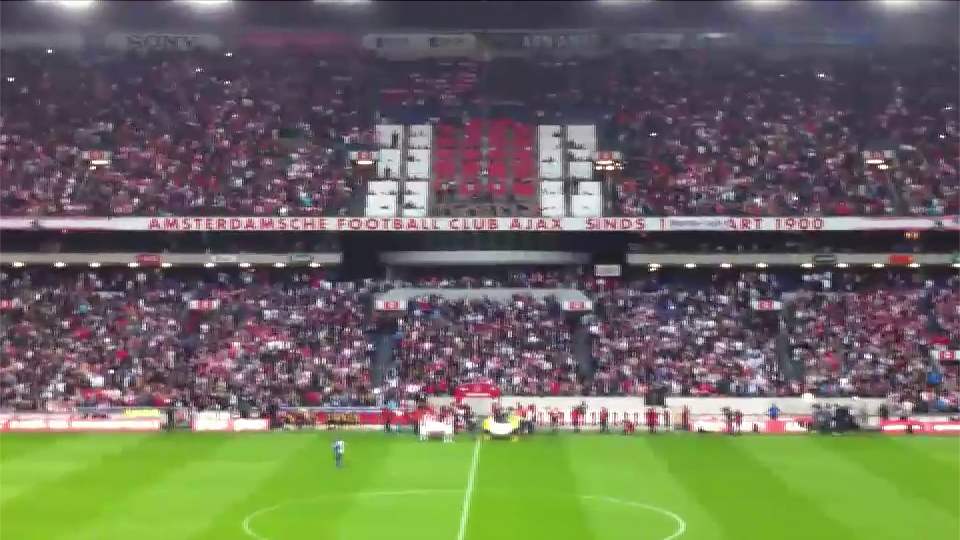}} \quad
    \subfloat[Predicted = 2.85, GT = 2.86]{\includegraphics[width=0.45\linewidth]{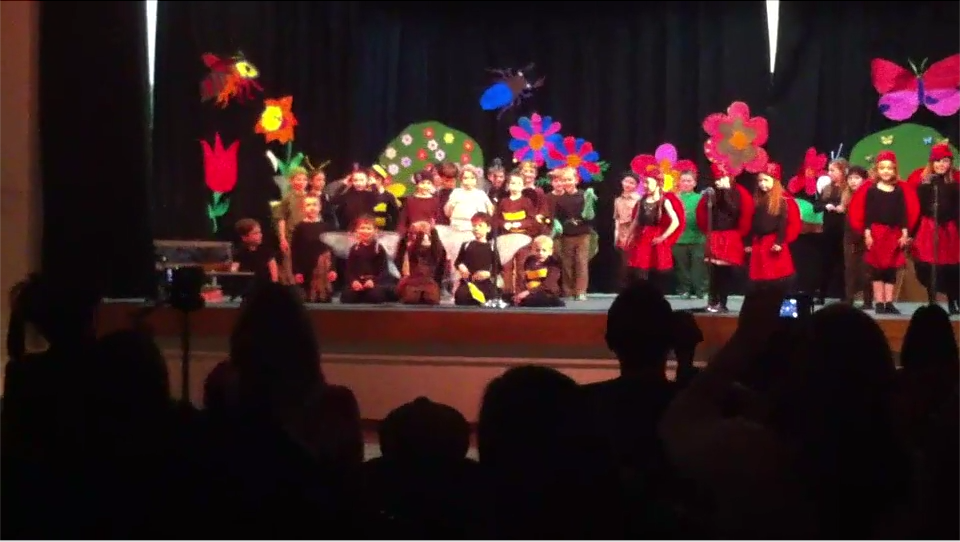}} \\
    \subfloat[Predicted = 3.33, GT = 2.52\label{fig:konvid_fail}]{\includegraphics[width=0.45\linewidth]{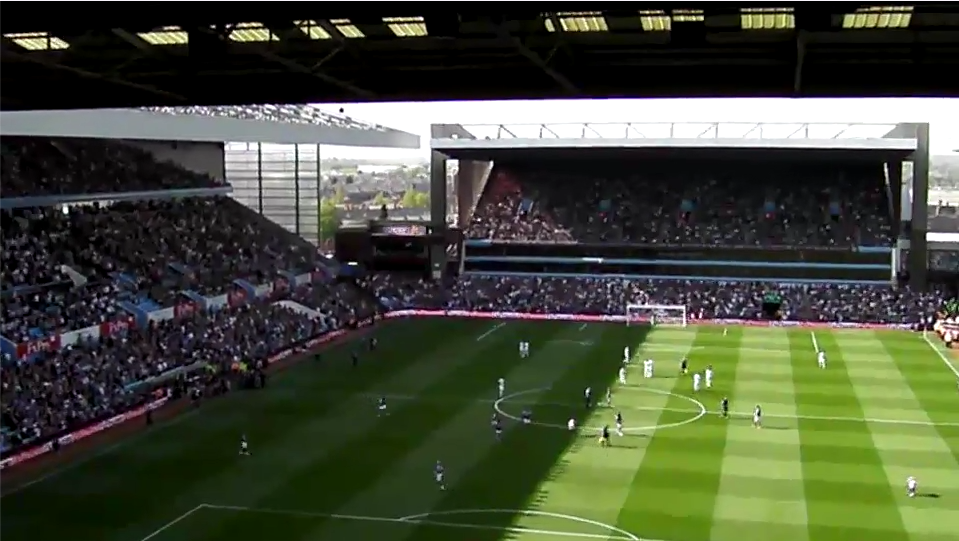}} \quad
    \subfloat[Predicted = 3.4, GT = 2.58]{\includegraphics[width=0.45\linewidth]{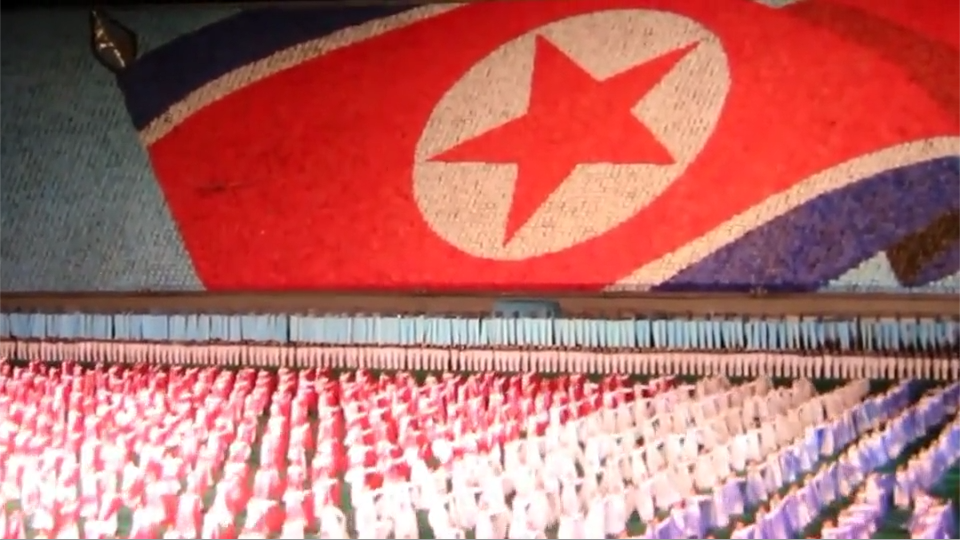}}
    \cprotect \caption{Illustration of successful and failure examples sampled from the KoNViD \cite{hosu2017konstanz} dataset. \textbf{Top-row}: Successful cases sampled from the \verb+7136906983.mp4+ and \verb+5668795950.mp4+ sequences. \textbf{Bottom-row}: Failure examples sampled from the \verb+5621374631.mp4+ and \verb+9524637688.mp4+ sequences. GT refers to ground truth quality scores.}
    \label{fig:authentic_qualitative}
\end{figure}

\begin{figure}[t]
    \centering
    \subfloat[Predicted = 45.1, GT = 46.8]{\includegraphics[width=0.45\linewidth]{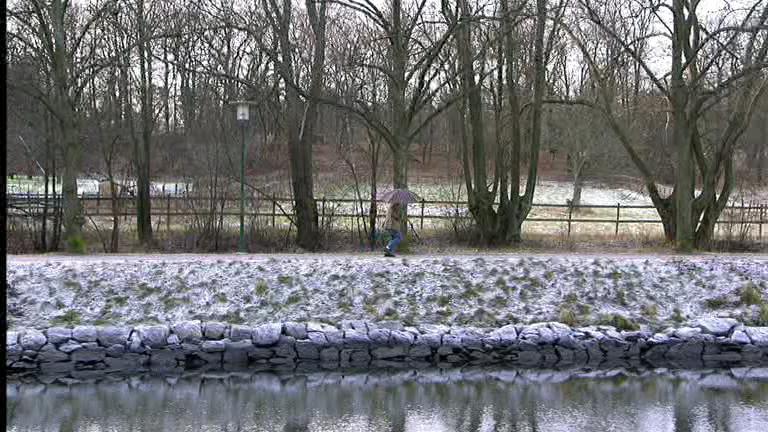}} \quad
    \subfloat[Predicted = 47.2, GT = 48]{\includegraphics[width=0.45\linewidth]{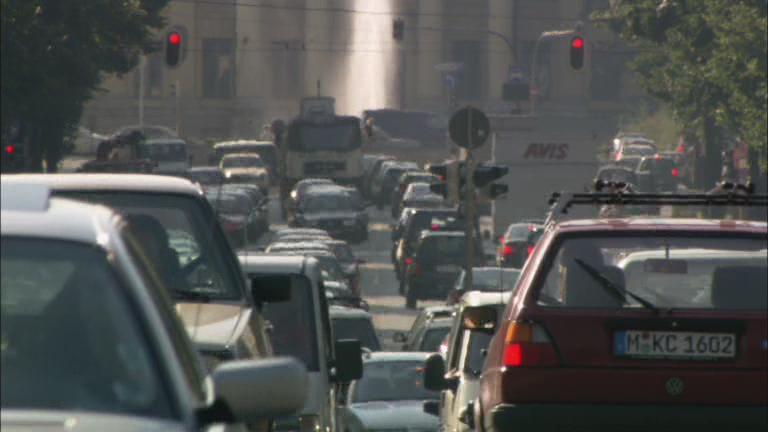}} \\
    \subfloat[Predicted = 55, GT = 68\label{fig:LIVE_fail_ip}]{\includegraphics[width=0.45\linewidth]{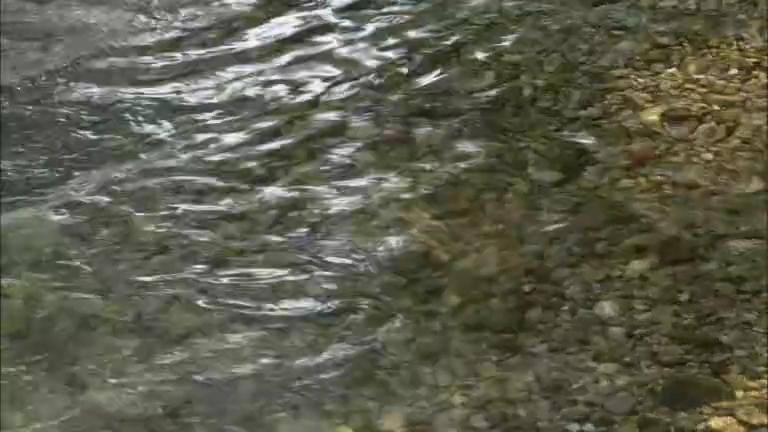}} \quad
    \subfloat[Predicted = 59.76, GT = 78.34\label{fig:LIVE_fail_wireless}]{\includegraphics[width=0.45\linewidth]{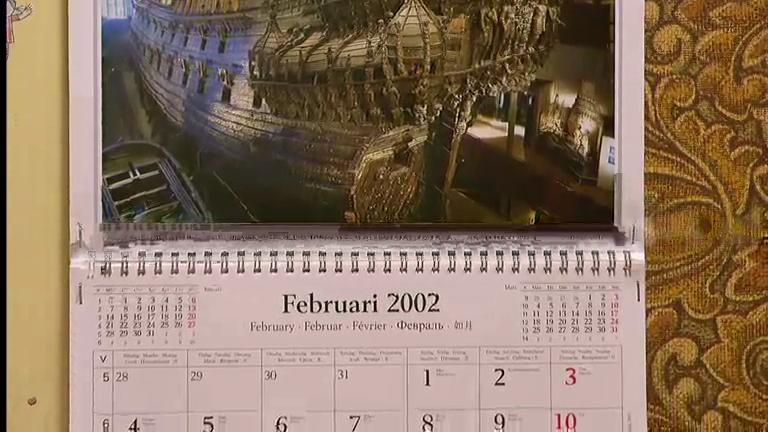}}
    \cprotect \caption{Illustration of successful and failure examples sampled from the LIVE-VQA \cite{seshadrinathan2010study} dataset. \textbf{Top-row}: Successful cases sampled from the \verb+pr14_25fps+ and \verb+rh13_25fps+ sequences. \textbf{Bottom-row}: Failure examples sampled from the \verb+rb6_25fps+ and \verb+mc2_50fps+ sequences. GT refers to ground truth quality scores.}
    \label{fig:synthetic_qualitative}
\end{figure}

In this subsection we qualitatively compare successful and failure examples of CONVIQT sampled from different VQA datasets in Fig. \ref{fig:authentic_qualitative} and Fig. \ref{fig:synthetic_qualitative}. Fig. \ref{fig:authentic_qualitative} contains cases with authentic distortions sampled from the KoNViD \cite{hosu2017konstanz} dataset, while Fig. \ref{fig:synthetic_qualitative} shows examples from LIVE-VQA \cite{seshadrinathan2010study} containing synthetic distortions. The failure example in Fig. \ref{fig:konvid_fail} contains flicker artifacts, while those in Figs. \ref{fig:LIVE_fail_ip} and \ref{fig:LIVE_fail_wireless} are corrupted by IP and wireless distortions respectively. We believe that the limited number of samples corresponding to these distortion types in the pretraining data were a contributing factor for these less accurate predictions.

\subsection{Limitations}
From Table \ref{table:synthetic_VQA} it may be observed that CONVIQT underperformed on the legacy datasets LIVE-VQA and CSIQ-VQA. In this subsection we investigate this behavior by analyzing the performance of CONVIQT on each individual distortion types present in these datasets.

These artifacts were not included in the pretraining data since they are less prevalent in current video sharing and streaming scenarios. Table \ref{table:LIVE_VQA_individual} and \ref{table:CSIQ_VQA_individual} contain the computed correlations of CONVIQT for each type of distortion present in LIVE-VQA and CSIQ-VQA respectively. It maybe observed from the Tables that CONVIQT obtained lower correlations on the wireless, IP and packet loss distortions than on the others. The lack of training examples corresponding to these distortion types could be a factor contributing to this lower performance.

\section{Conclusion and Future Work}
\label{sec:conclusion}

We proposed a self-supervised training procedure to obtain perceptually relevant video quality representations. The model was trained with the goal of discriminating different distortion types and degradation degrees using a large unlabeled video database, containing videos afflicted by both synthetic and realistic distortions. We performed comprehensive evaluations of our proposed design, and found that the performance of CONVIQT is comparable to many supervised deep VQA models. Notably, CONVIQT achieves this performance even without fine-tuning, underscoring the effectiveness of the auxiliary task. We conducted ablation experiments to understand the significance of the temporal transformations, and inferred that these transformations lead to better generalization performance. We also studied the impact of the distortion types present in training data, and deduced that a training database containing a combination of synthetic and authentic distortions yielded more robust features. A software release of CONVIQT has been released online\footnote{\url{https://github.com/pavancm/CONVIQT}} to promote reproducible research.

Though CONVIQT achieves competitive performance on VFR videos, there is significant room for improvement, \textit{viz.}, the representations can be made more sensitive to reflect perceptual quality changes occurring due to frame rate variations. As part of future work, we would like to explore auxiliary tasks which promote feature learning with high sensitivity to frame rate artifacts.

\begin{table}[t]
\caption{CONVIQT performance comparison on individual distortion types present in the LIVE-VQA dataset.}
\label{table:LIVE_VQA_individual}
    \centering
    \footnotesize
    \begin{tabular}{|c||c|c|}
        \hline
        Distortion Type & SROCC$\uparrow$ & PLCC$\uparrow$ \\ \hline \hline
        Wireless & 0.595 & 0.629 \\ 
        IP & 0.486 & 0.510 \\ 
        H264 & 0.738 & 0.719 \\ 
        MPEG & 0.810 & 0.800 \\ \hline
        Overall & \textbf{0.622} & \textbf{0.595} \\
        \hline
    \end{tabular}
\end{table}

\section{Acknowledgment}
This work was supported by grant number 2019844 for the National Science Foundation AI Institute for Foundations of Machine Learning (IFML). The authors would also like to acknowledge the Texas Advanced Computing Center (TACC) for providing computational resources that contributed to this research, and YouTube for supporting this work. 

\begin{table}[t]
\caption{CONVIQT performance comparison on individual distortion types present in the CSIQ-VQA dataset.}
\label{table:CSIQ_VQA_individual}
    \centering
    \footnotesize
    \begin{tabular}{|c||c|c|}
        \hline
        Distortion Type & SROCC$\uparrow$ & PLCC$\uparrow$ \\ \hline \hline
        H264 & 0.817 & 0.819 \\ 
        Packet Loss & 0.533 & 0.417 \\ 
        MJPEG & 0.800 & 0.806 \\ 
        Wavelet & 0.867 & 0.867 \\ 
        White Noise & 0.800 & 0.796 \\
        HEVC & 0.717 & 0.735 \\ \hline
        Overall & \textbf{0.766} & \textbf{0.749} \\
        \hline
    \end{tabular}
\end{table}

\bibliographystyle{IEEEtran}
\bibliography{references}

\end{document}